\documentclass{aa}
\usepackage{graphicx}
\usepackage[varg]{txfonts}
\usepackage{natbib}
\usepackage{xcolor}

\newcommand{\zav}[1]{\left(#1\right)}
\newcommand{\hzav}[1]{\left[#1\right]}
\newcommand{\azav}[1]{\left|#1\right|}

\newlength\staretab

\newcommand\de{\text{d}}

\newcommand\kms{\ensuremath{\text{km}\,\text{s}^{-1}}}

\newcommand\rk{\ensuremath{r_\text{K}}}

\begin{document}

\title{Photometric signatures of corotating magnetospheres of hot stars governed
by higher-order magnetic multipoles}
\titlerunning{Photometric signatures of corotating magnetospheres governed
by higher-order multipoles}

\author{J.~Krti\v cka$^1$ \and Z.~Mikul\'a\v sek$^1$ \and P.~Kurf\"urst$^1$ \and
M.~E.~Oksala$^{2,3}$}
\authorrunning{J.~Krti\v cka et al.}

\institute{Department of Theoretical Physics and Astrophysics,
           Masaryk University, Kotl\'a\v rsk\' a 2, CZ-611\,37 Brno, Czech
           Republic \and
           Department of Physics, California Lutheran University, 60 West Olsen
           Road \#3700, Thousand Oaks, CA 91360, USA \and
           LESIA, Paris Observatory, PSL University, CNRS, Sorbonne University,
           Univ. Paris Diderot, Sorbonne Paris Cit, 5 place Jules 14 Janssen,
           92195 Meudon, France
}

\date{Received}

\abstract{The light curves of magnetic, chemically peculiar stars typically show
periodic variability due to surface spots that in most cases can be modeled by
low-order harmonic expansion. However, high-precision satellite photometry
reveals tiny complex features in the light curves of some of these stars that
are difficult to explain as caused by a surface phenomenon under reasonable
assumptions. These features might originate from light extinction in corotating
magnetospheric clouds supported by a complex magnetic field dominated by
higher-order multipoles.}{We aim to understand the photometric signatures of
corotating magnetospheres that are governed by higher-order multipoles.}{We
determined the location of magnetospheric clouds from the minima of the
effective potential along the magnetic field lines for different orders of
multipoles and their combination. From the derived magnetospheric density
distribution, we calculated light curves accounting for absorption and
subsequent emission of light.} {For axisymmetric multipoles, the rigidly
rotating magnetosphere model is able to explain the observed tiny features in
the light curves only when the higher-order multipoles dominate the magnetic
field not only at the stellar surface, but even at the Kepler radius. However,
even a relatively weak nonaxisymmetric component leads to warping of equilibrium
surfaces. This introduces structures that can explain the tiny features observed
in the light curves of chemically peculiar stars. The light emission contributes
to the light variability only if a significant fraction of light is absorbed in
the magnetosphere.}{}


\maketitle

\section{Introduction}

Classical chemically peculiar stars are stars in the upper part of the main sequence,
where the diffusion  due to competing radiative and gravitational forces leads
to chemical peculiarity \citep{michzak,vadog,vimiri}. In chemically peculiar
stars, the elements concentrate in vast surface spots (patches), which are
formed by the influence of the magnetic field \citep{alestimag} and likely also
by other processes \citep{putting,jagnemag}. As a result of rotational
modulation, the spots cause periodic spectroscopic and 
photometric variability.

Unlike the spots on cool stars, the patches on chemically peculiar stars are
mostly stable over at least decades \citep[e.g.,][]{adelct,brzda} and have the
same effective temperature as the rest of the surface. The latter follows from
the fact that most of the light variability of these stars can be explained as a
result of flux redistribution due to bound-free \citep{peter,myhd37776} and
bound-bound transitions \citep{vlci,molnar,myhr7224,seuma,prvalis} of helium and
heavier elements such as silicon, chromium, and iron.

It is commonly assumed that as a result of their modulation by rotation, light
curves of chemically peculiar stars can be typically reproduced using low-order
harmonic expansion \citep[e.g.,][]{mikzoo,zrychrot}. However, high-precision
satellite photometry changed this picture. In addition to the variability that can be
attributed to spots, a large fraction of hot chemically peculiar stars
(Mikul\'a\v sek et al., in preparation) shows tiny complex features (also known as~warps)
on their light curves \citep{humkep,mikland}. These narrow features in the light
curves do not appear on theoretically simulated light curves of chemically
peculiar stars \citep[e.g.,][]{myhr7224,seuma}. Moreover, narrow features are
difficult to explain by surface modulation under reasonable assumptions
about the intensity contrast of spots and their sharpness \citep[Sects. 5.3 and 5.5]{milanphd}.

On the other hand, the rotational light
variability of chemically peculiar stars can be caused by sources other than surface patches. A large
fraction of chemically peculiar stars may have winds driven by the line radiative
force \citep{abbobla,babelb,metuje}. Even below the limit of hydrogen-dominated winds, however, purely metallic winds are possible \citep{babela}. Within
the rigidly rotating magnetosphere model, the stellar wind flows along magnetic
field lines and accumulates in the region of minimum effective potential given
by centrifugal and gravitational forces. The magnetospheric clouds established
by this process occult part of the stellar surface, which may lead to dips 
(eclipses) in the light curve \citep{labor,nakaji,smigro,towog}. Only a few stars show dips on their light curves in ground-based
observations \citep[e.g.,][]{towog,druhasigorie,grunzak}. On the other
hand, many more such stars may be detected with high-precision satellite
photometry, which can be connected with the absorption of stellar radiation by
corotating magnetospheric clouds.

In principle, the observed tiny features in the light curves can be also
interpreted in terms of additional emission (bumps) instead of absorption.
However, the width of these features would imply a strong beaming of the radiation,
which might be less conceivable than simple absorption by intervening
clouds.

The accumulation of matter in dipolar field with a high obliquity leads to two
isolated minima in the light curve \citep{towog}. Therefore, a magnetic field
dominated by higher-order multipoles might explain the light curves
with numerous warps. A similar model was proposed by \citet{zahradajakomy} to
provide prominence support in the magnetospheres of cool stars. We examine this
possibility in this paper, which is organized as follows.
Section~\ref{rotmagmulti} gives the structure of equilibrium surfaces at which
the magnetospheric matter accumulates for a magnetic field dominated by higher-order multipoles, while Sect.~\ref{krivulemulti} predicts the light curves induced
by these magnetospheres. Finally, in Sect.~\ref{dalsi} we discuss an application
of our model to typical chemically peculiar stars with warped light curves, and
Sect.~\ref{zaver} presents our conclusions. 

\section{Rotating magnetosphere model for general multipoles}
\label{rotmagmulti}

\begin{figure*}
\includegraphics[width=0.5\textwidth]{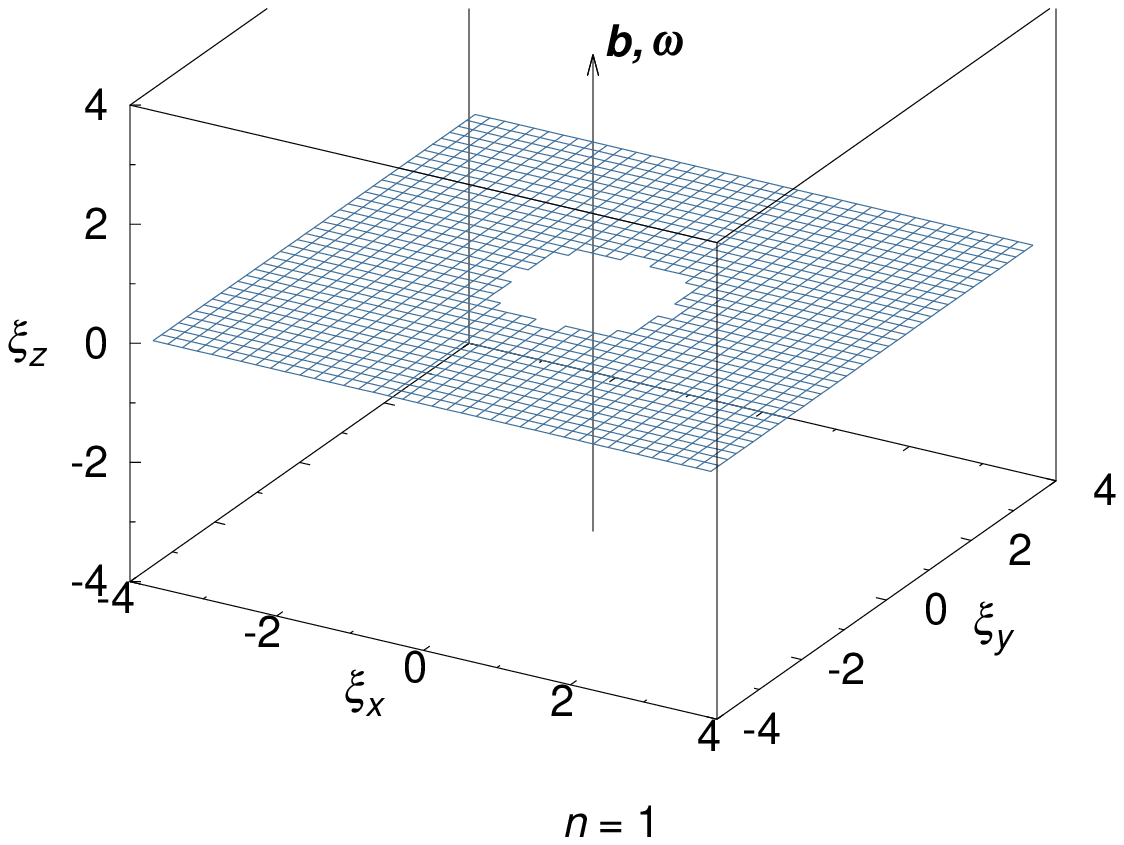}
\includegraphics[width=0.5\textwidth]{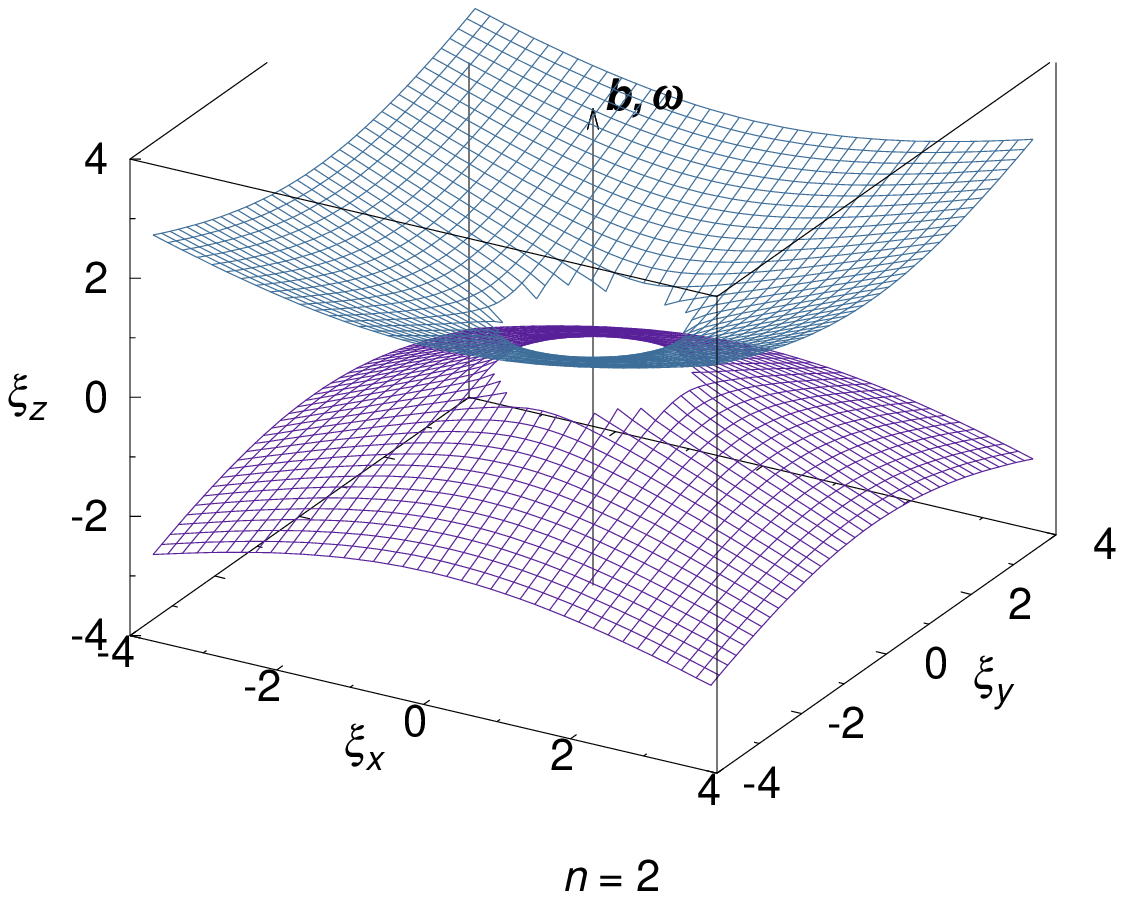}
\includegraphics[width=0.5\textwidth]{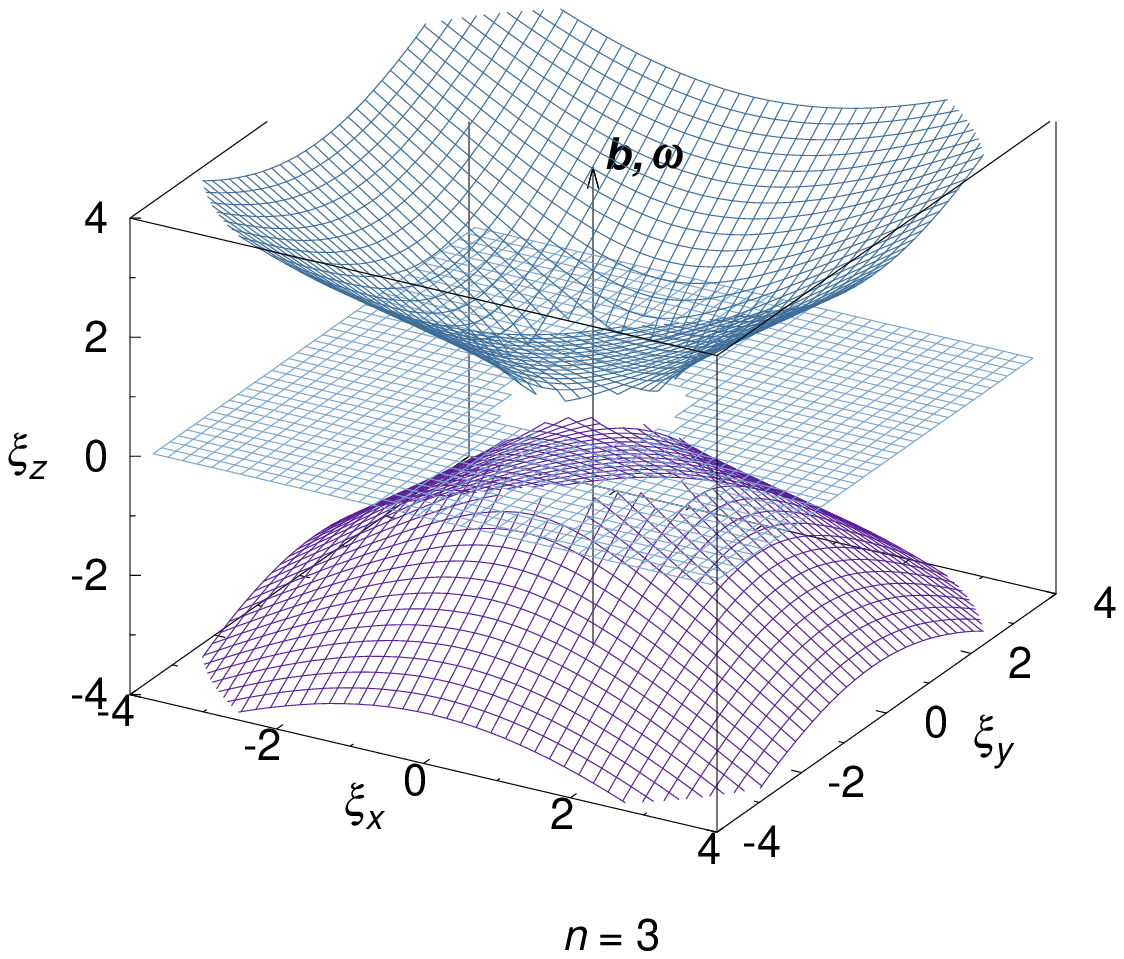}
\includegraphics[width=0.5\textwidth]{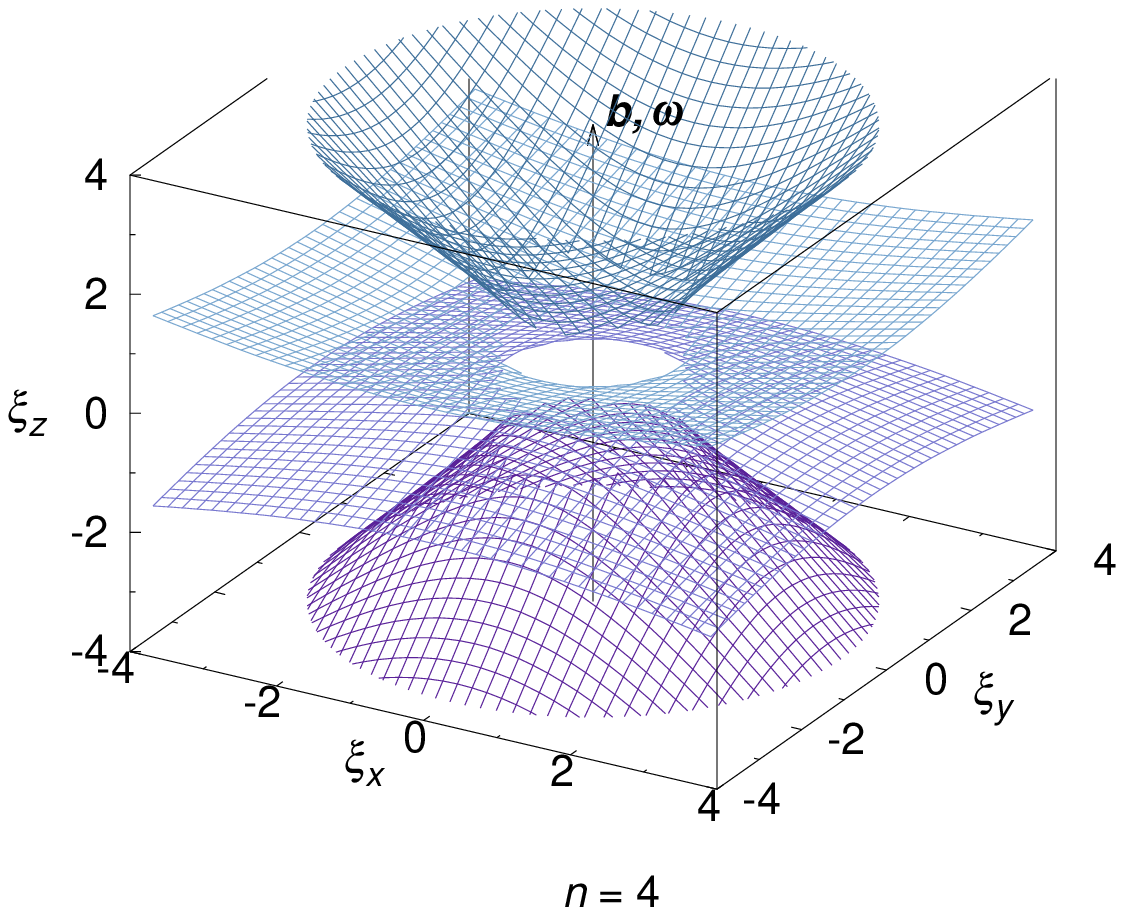}
\caption{Equilibrium surfaces Eq.~\eqref{rovno} fulfilling the stability
condition Eq.~\eqref{stab} for a field aligned with the stellar rotational axis
and multipoles with different orders.} 
\label{mrakbrovno}
\end{figure*}

A physically motivated distribution of the matter in the magnetosphere of a rotating
star can be obtained by elaborating the models proposed by \citet{prus} and
\citet{towo}. The latter model is more general and allows determining the
density distribution accounting for the wind mass-loss rate modulation due to
the magnetic field. The model was successfully applied not only to study
the light curves due to circumstellar absorption, but also to provide a detailed
interpretation of H$\alpha$ magnetospheric emission
\citep[e.g.,][]{mysigorie,shalfa} and polarimetry \citep{carsigopol}.

In a corotating circumstellar magnetosphere, the ionized matter is at rest when
the sum of the gravitational and centrifugal force (per unit of mass)
$\boldsymbol{f}$ is perpendicular to the magnetic field,
\begin{equation}
\label{rovno}
\boldsymbol{f}\cdot\boldsymbol{b}=0,
\end{equation}
where
\begin{equation}
\label{efko}
\boldsymbol{f}=\boldsymbol{g}-\boldsymbol{\Omega}\times
\zav{\boldsymbol{\Omega}\times\boldsymbol{r}}=
\frac{GM}{r_\text{K}^2}\hzav{\boldsymbol{\xi}\zav{1-\frac{1}{\xi^3}}-
\boldsymbol{\omega}\zav{\boldsymbol{\omega}\cdot\boldsymbol{\xi}}},
\end{equation}
and $\boldsymbol{g}$ stands for gravity acceleration,
$\boldsymbol{\Omega}=\Omega\boldsymbol{\omega}$ is the angular
frequency, $\boldsymbol{b}=\boldsymbol{B}/B$ is the unit vector in the direction
of magnetic field $\boldsymbol{B}$, and $\boldsymbol{\xi}=\boldsymbol{r}/\rk$ is
the radius vector $\boldsymbol{r}$ in units of the Kepler radius,
\begin{equation}
\rk=\zav{\frac{GM}{\Omega^2}}^{1/3}.
\end{equation}
The stability condition further requires that the displacement of an element
from the equilibrium position
along the field line $\boldsymbol{b}$ leads to the appearance of the force in
the opposite direction, that is, 
\begin{equation}
\label{stab}
f'\equiv\zav{\boldsymbol{b}\nabla}\boldsymbol{f}\cdot\boldsymbol{b}+
\boldsymbol{f} \cdot\zav{\boldsymbol{b}\nabla}\boldsymbol{b}<0.
\end{equation}
In the model of \citet{towo}, this corresponds to the potential minimum along
the field line. Inserting the effective force in Eq.~\eqref{efko}, the stability
condition Eq.~\eqref{stab} can be rewritten as
\begin{equation}
\label{babice}
\frac{1}{\xi^3}\hzav{\frac{3}{\xi^2}\zav{\boldsymbol{b}\cdot\boldsymbol{\xi}}^2-1}+
1-\zav{\boldsymbol{\omega}\cdot\boldsymbol{b}}^2+
\boldsymbol{f}_\xi \cdot\zav{\boldsymbol{b}\nabla_\xi}\boldsymbol{b}<0.
\end{equation}
Here $\nabla_\xi$ denotes the gradient in nondimensional units,
$\nabla_\xi=r_\text{K}\nabla$ and
$\boldsymbol{f}_\xi=\boldsymbol{f}/(GM/r_\text{K}^2)$.

\subsection{Field aligned with rotational axis}
\label{fiali}

\begin{figure*}
\includegraphics[width=0.5\textwidth]{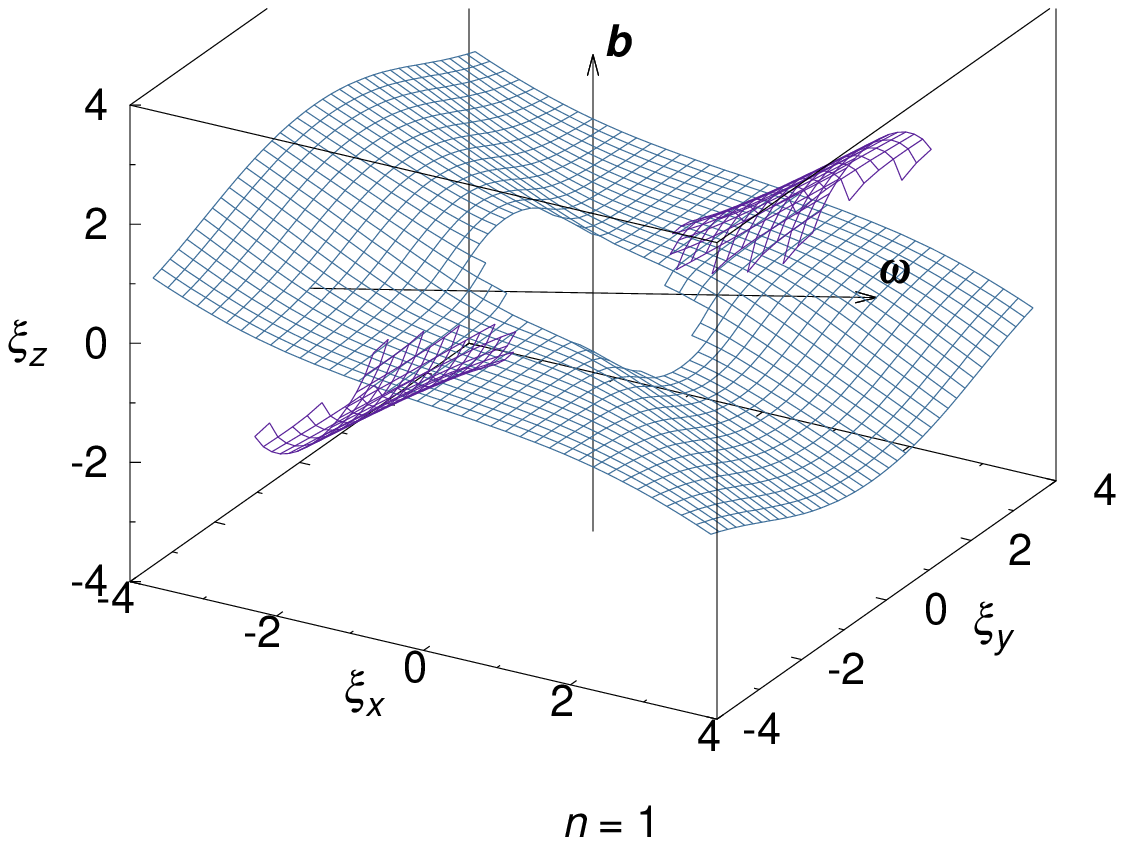}
\includegraphics[width=0.5\textwidth]{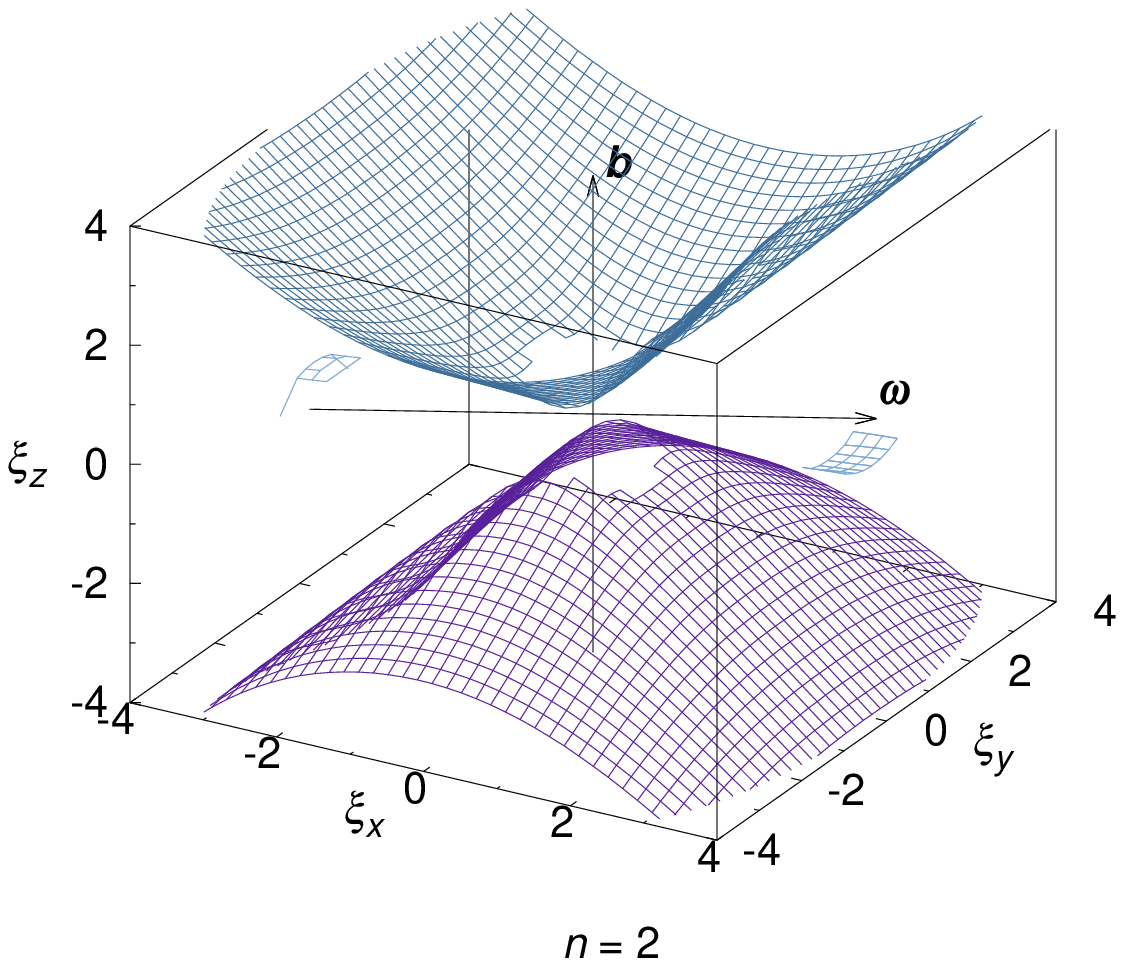}
\includegraphics[width=0.5\textwidth]{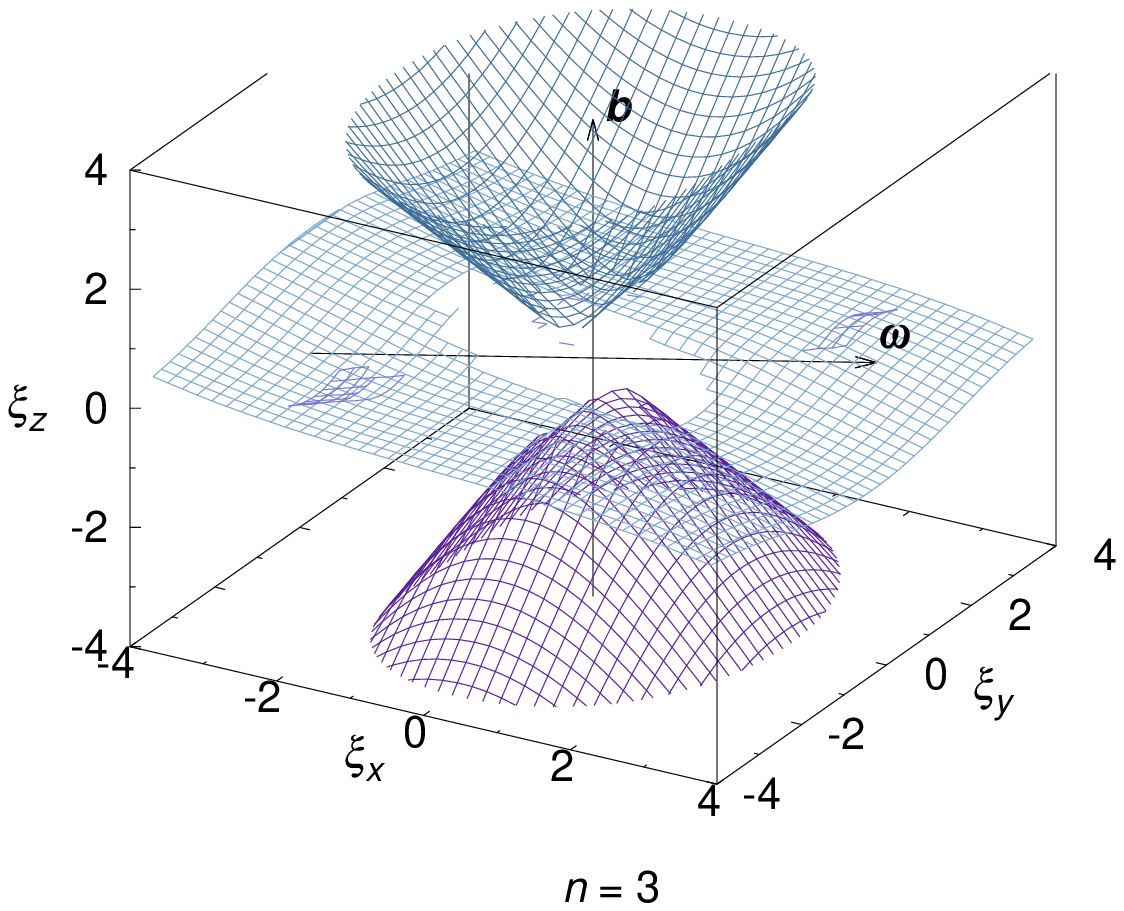}
\includegraphics[width=0.5\textwidth]{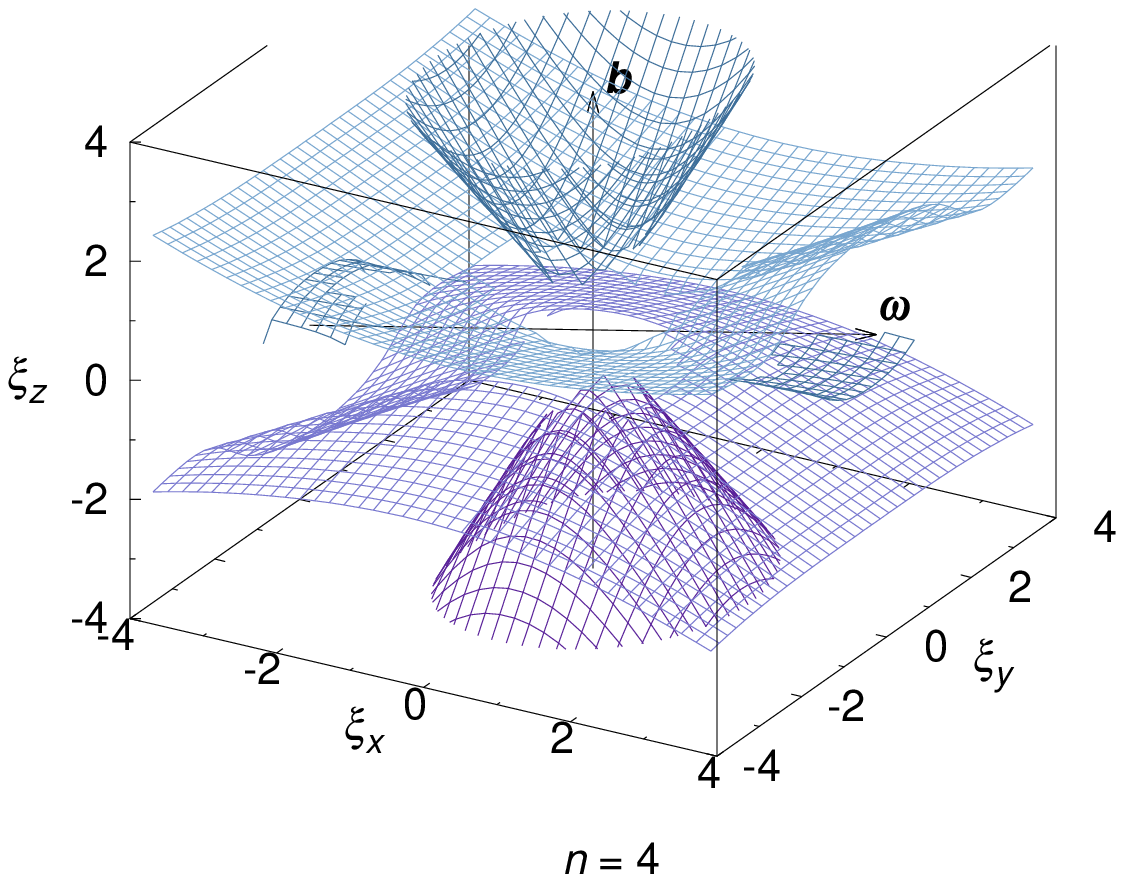}
\caption{Equilibrium surfaces Eq.~\eqref{rovno} fulfilling the stability
condition Eq.~\eqref{stab} for a rotational axis $\boldsymbol{\omega}$ tilted by
$5/12\,\pi$ with respect to the axis of the magnetic field $\boldsymbol{b}$ for
multipoles with different orders. Both axes are denoted in the graph.} 
\label{mrakpsi75}
\end{figure*}

In the simplest case, the axes of the multipole and the rotational axis are
parallel. The magnetic field strength is modeled using co-aligned
axisymmetric multipoles of the order $n,$
\begin{equation}
\label{multipol}
\boldsymbol{B}=\frac{B_n}{r^{n+2}}P_n(\cos\theta)\,\boldsymbol{e}_r-
\frac{1}{n+1}\frac{B_n}{r^{n+2}}P_n^1(\cos\theta)\,\boldsymbol{e}_\theta,
\end{equation}
where $P_n$ and $P_n^1$ denote the Legendre and associated Legendre polynomials,
respectively. We do not account for a field with a nonzero azimuthal component in
the initial calculations. The plots of the equilibrium surfaces Eq.~\eqref{rovno}
fulfilling the stability condition Eq.~\eqref{stab} are given in
Fig.~\ref{mrakbrovno}.

For a dipolar case ($n=1$), the resulting equilibrium surface corresponds to
an equatorial plane with a hole. In the equatorial plane, the force
$\boldsymbol{f}$ is perpendicular to the magnetic field fulfilling the
equilibrium condition of Eq.~\eqref{rovno}. The central hole appears from the stability
condition Eq.~\eqref{stab}. For a dipolar field in the equatorial plane,
\mbox{$\boldsymbol{b}\cdot\boldsymbol{\xi}=0$} and
$\boldsymbol{\omega}\cdot\boldsymbol{b}=1$, and after an evaluation of the
gradient term, the stability condition consequently is \citep{prus,towo}
\begin{equation}
\label{blansko}
\xi>\zav{\frac{2}{3}}^{1/3}.
\end{equation}
Therefore, as a result of the influence of the magnetic field, the matter is
stable even slightly below the Kepler radius $\xi=1$.

There are additional equilibrium chimney-like surfaces in the polar direction
\citep{prus}, where the force $\boldsymbol{f}$ is perpendicular to the magnetic
field. However, the equilibrium on these surfaces is unstable because either
gravity or centrifugal force drives the matter out of the equilibrium position.

The quadrupole ($n=2$) has two lobes below and above the equatorial plane,
therefore there are two surfaces at which the matter is in equilibrium and is
stable (Fig.~\ref{mrakbrovno}; see also \citealt{zahradajakomy}). In
general, there is an equilibrium surface for each lobe of the multipole. The
apex of the lobe appears where the radial component of the magnetic field is
zero, which from Eq.~\eqref{multipol} corresponds to the root of the appropriate
Legendre polynomial. The number of roots is given by an order of multipole,
therefore the number of lobes is equal to the order of the multipole. The
surfaces show a mirror symmetry with respect to the equatorial plane, and
this plane consequently corresponds to stable solutions for odd multipoles. At large
distances from the star, the effective force in the equilibrium condition
$\boldsymbol{f}\cdot\boldsymbol{b}=0$ is dominated by the centrifugal force,
which is perpendicular to the rotational axis. Therefore, the equilibrium
condition is fulfilled at the surface where the magnetic field is parallel to
the rotational axis. Individual magnetic field lines are similar, therefore
the magnetic field is parallel to the rotational axis for a particular $\theta$.
As a result, at large distances from the star, the equilibrium surfaces is
conical.

This also explains why there is one equilibrium surface for each order of the
multipole. The order of multipole determines the number of lobes. In each lobe,
there is just one magnetic latitude where the magnetic field lines are
perpendicular to the effective force, therefore there is just one equilibrium
surface per each lobe or multipolar order. In the equatorial plane, the effective
force is radial, therefore the equilibrium surface appears at the apex of the
lobe where the radial magnetic field component is zero. This is given by the
root of the corresponding Legendre polynomial. Outside the equatorial plane, the
effective force is no longer radial, therefore the equilibrium surface does
not appear at the apex of the lobe.

For odd multipoles, one of the equilibrium surfaces always appears in the
equatorial plane. Taking into account Eq.~\eqref{efko}, the stability
condition Eq.~\eqref{babice} simplifies to $-1/\xi^3+\gamma(1-1/\xi^3)<0$, where
$\gamma=\boldsymbol\xi\zav{\boldsymbol{b}\nabla_\xi}\boldsymbol{b}$. Evaluating
the directional derivative in the equatorial plane in spherical coordinates
taking into account the properties of the Legendre polynomials and Eq.~\eqref{multipol}
gives $\gamma=-n-2$. Therefore, the solution of the inequality leads to a generalized
condition of Eq.~\eqref{blansko}, 
\begin{equation}
\label{blanskolibn}
\xi>\zav{\frac{n+1}{n+2}}^{1/3}.
\end{equation}
This means that with an increasing order of the multipole, the radius of the central
hole grows from $\xi=(2/3)^{1/3}$ (Eq.~\eqref{blansko}) to $\xi=1$. This
corresponds to a change in equilibrium condition with the variation of the order of
the multipole. For a uniform magnetic field parallel with the rotational axis, the matter
would be stable anywhere in the equatorial plane. With increasing order of the
multipole, the magnetic field lines become increasingly bended. Therefore, for
multipoles with high order, the matter becomes stable if the centrifugal force is
stronger than gravity, as is usually the case without a magnetic field, and the radius
of the central hole approaches $\xi=1$. The numerical results show that below and
above the equatorial plane, the matter is stable outside the cylinder with radius
$\xi\approx1$.

%

\subsection{Misaligned rotation}

\begin{figure*}
\includegraphics[width=0.48\textwidth]{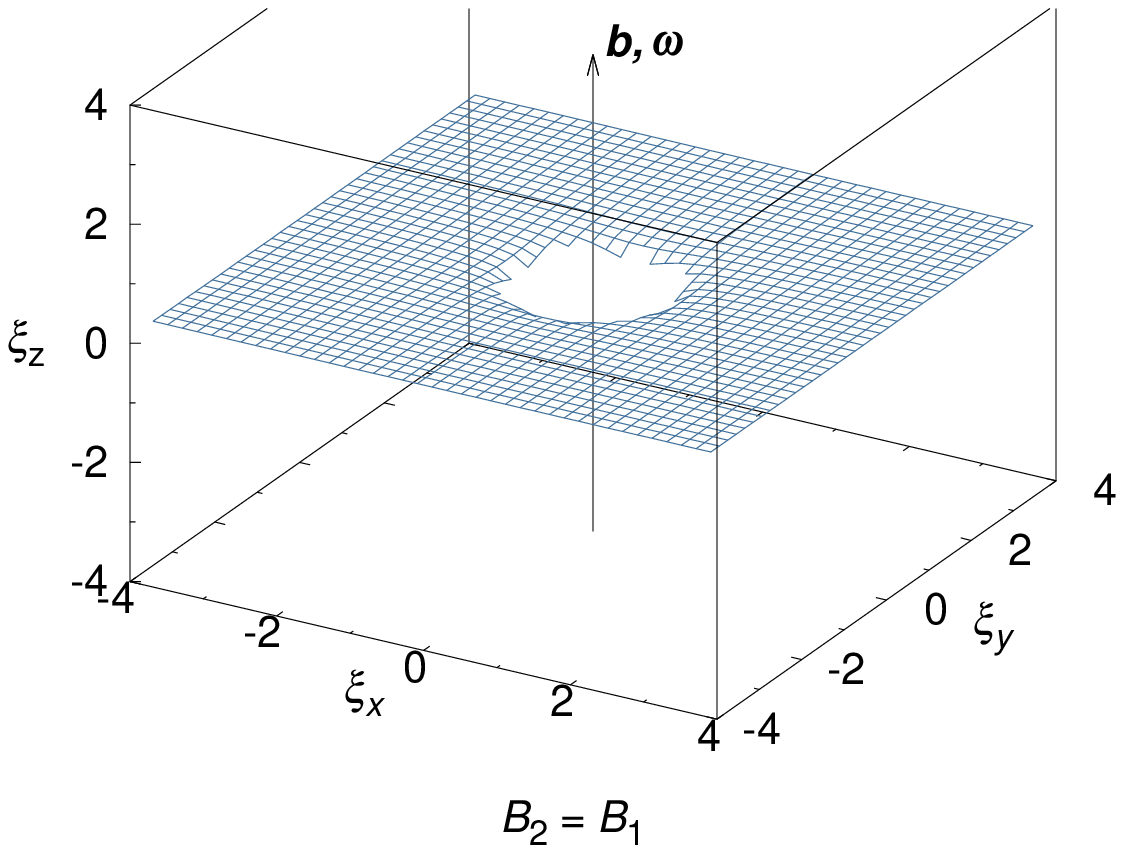}
\includegraphics[width=0.48\textwidth]{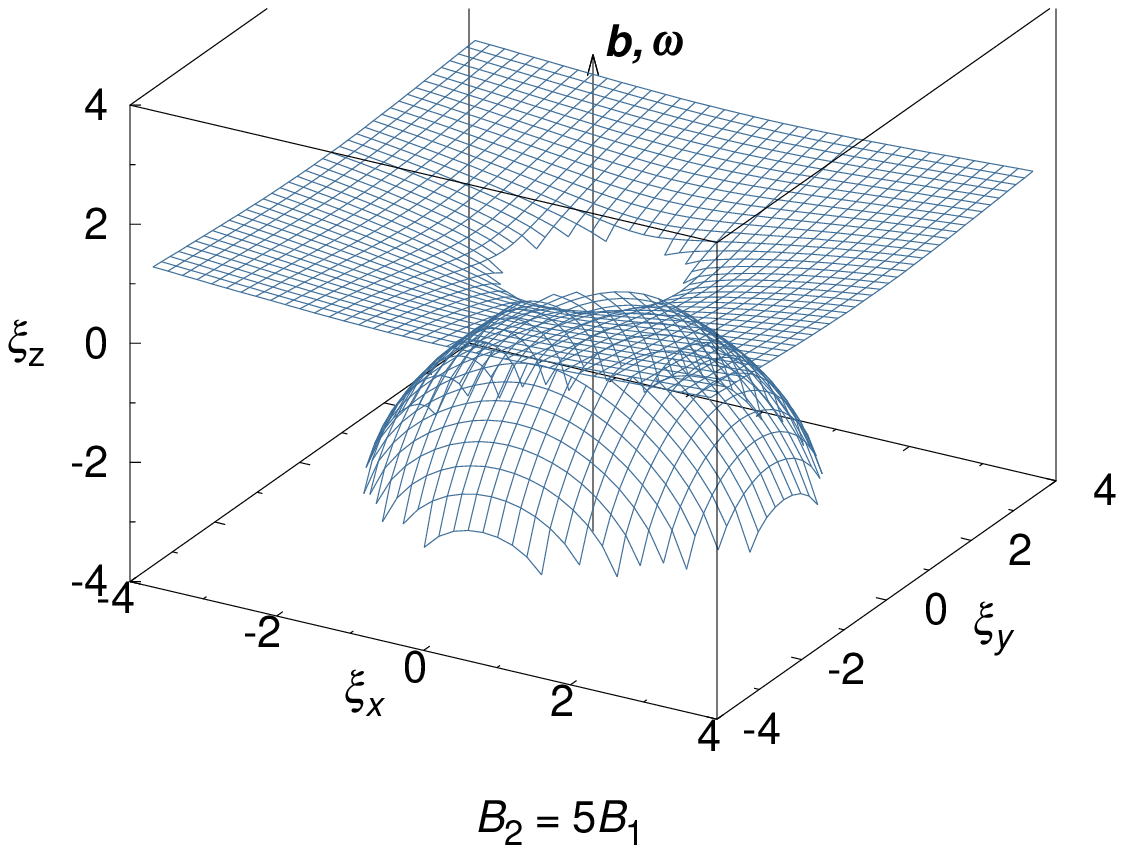}
\includegraphics[width=0.48\textwidth]{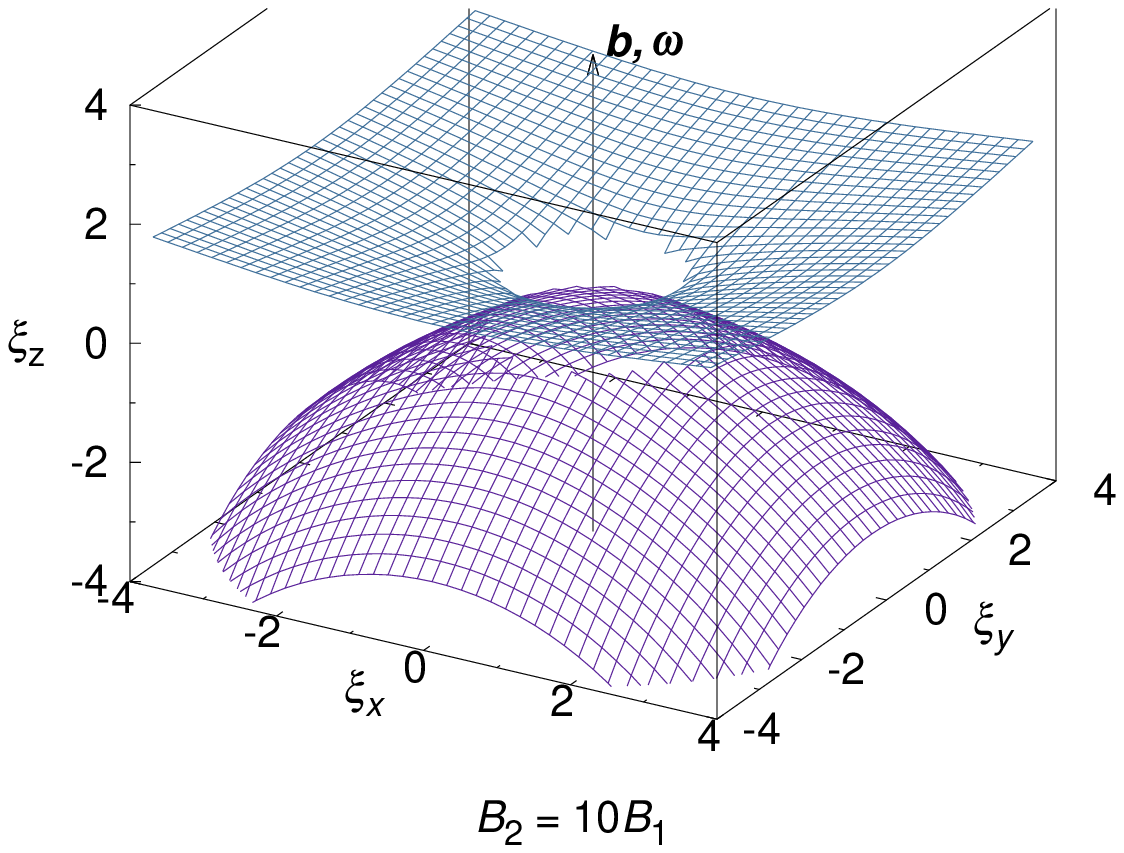}
\includegraphics[width=0.48\textwidth]{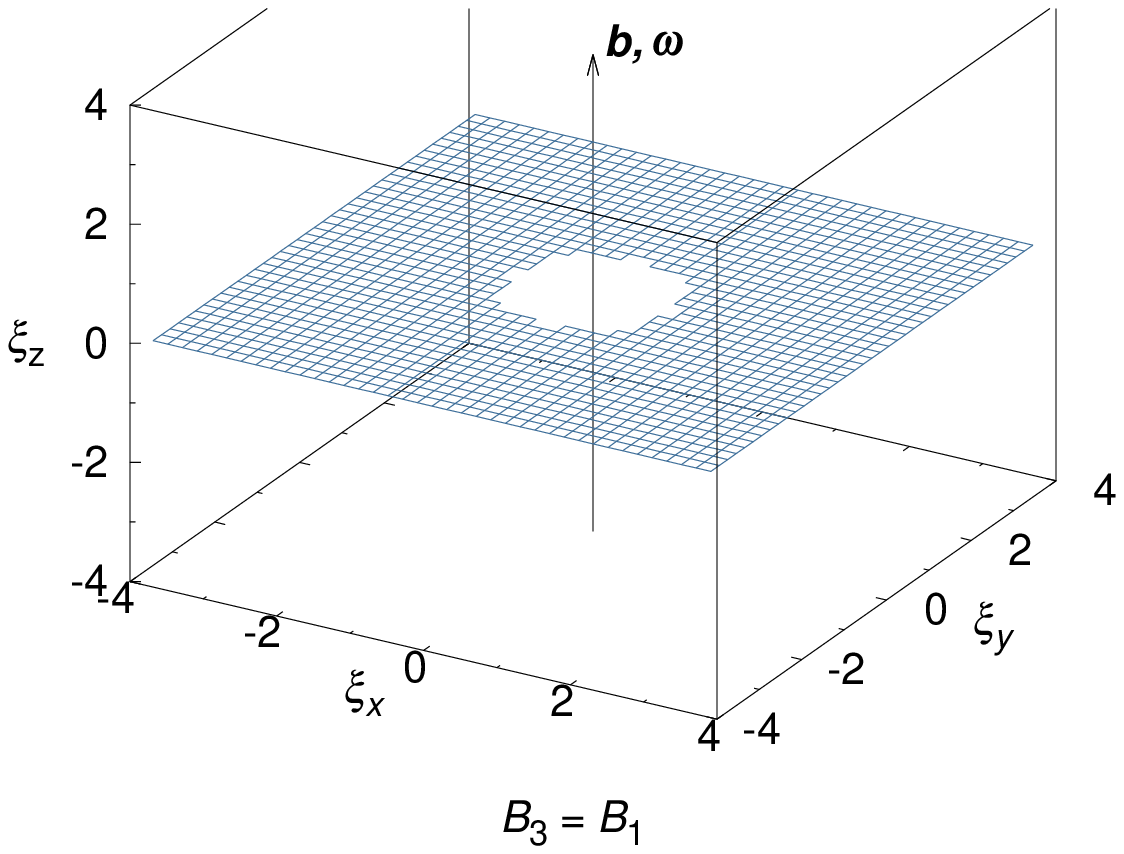}
\includegraphics[width=0.48\textwidth]{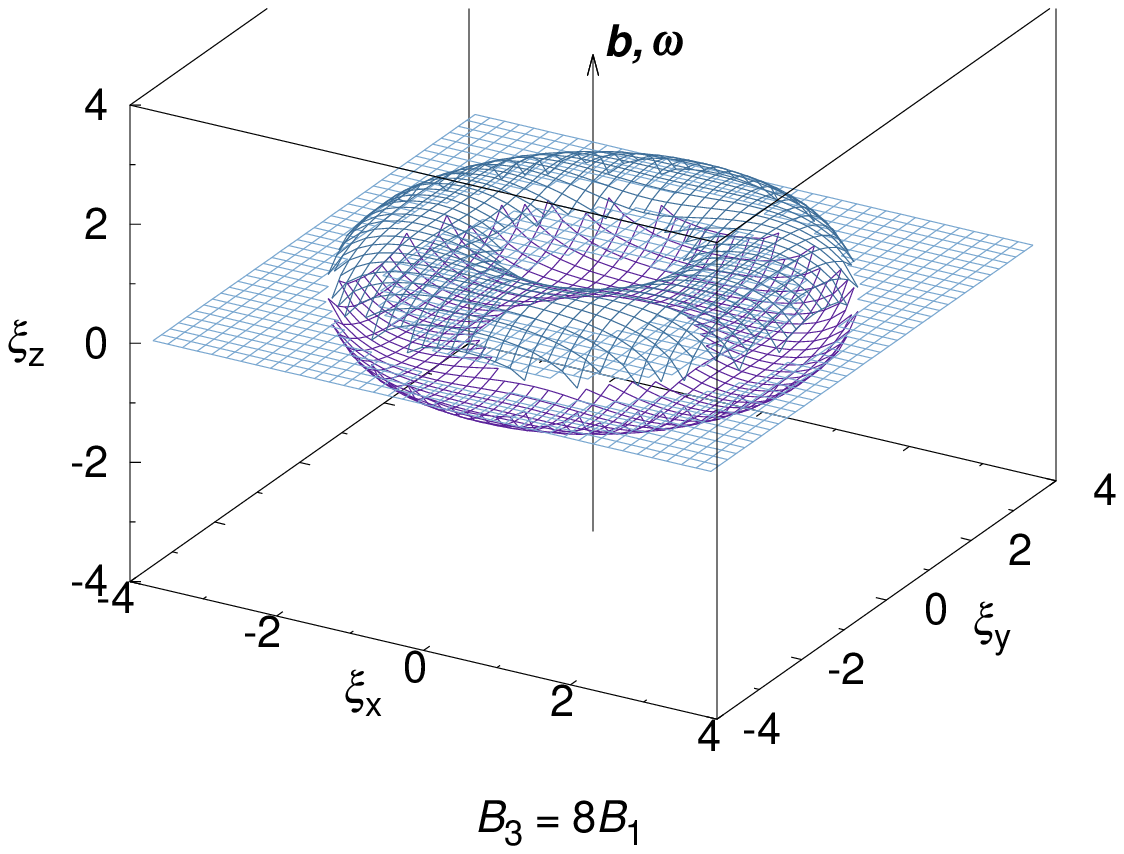}
\includegraphics[width=0.48\textwidth]{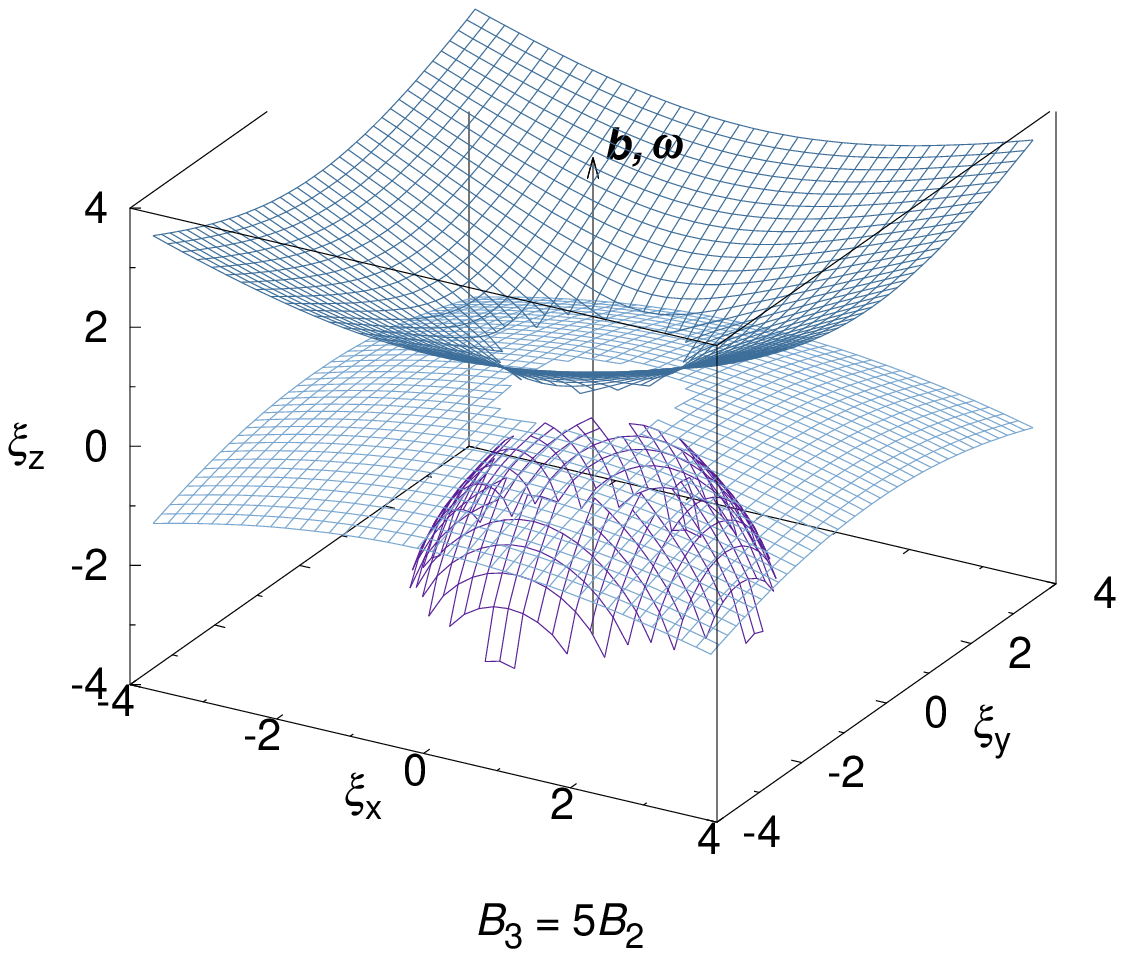}
\caption{Equilibrium surfaces fulfilling the stability condition for a field
aligned with the rotational axis and a combination of multipoles with different orders.}
\label{mraknm}
\end{figure*}

The axis of the magnetic field is typically tilted with respect to the
rotational axis in magnetic, hot stars. We denote the angle between these axes
(the magnetic field obliquity) by $\beta$. For $\beta>0,$ the equilibrium surfaces
become warped, as we show in Fig.~\ref{mrakpsi75}. As in aligned case, there is
one equilibrium surface per multipole order. In addition, part of the
chimney-like surface that was unstable for a field aligned with the rotational axis
becomes stable because potential minima appear at the surface.

In the case of misaligned rotation, the axis of the cylinder oriented along the
rotational axis with radius $\xi=1$, outside of which the stability
condition is fulfilled, is tilted with respect to the magnetic axis. As a
result, the region closest to the star where the matter is in stable equilibrium,
appears at the intersection of the magnetic and rotational equators \citep{towo}.

\subsection{General combination of coaligned multipoles}

\begin{figure}
\includegraphics[width=0.5\textwidth]{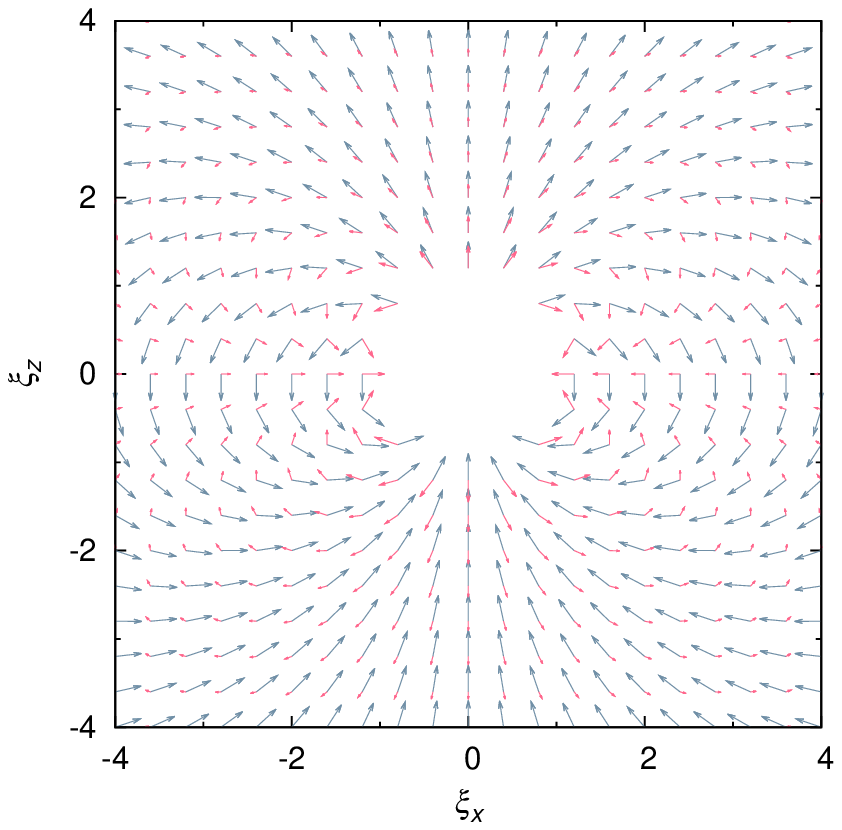}
\caption{Magnetic field vector for the dipolar (blue) and quadrupolar (red)
field plotted in the $y=0$ plane. The dipolar component is plotted with the unit
length vector, and the length of the quadrupolar field vector is reduced by a factor
of $1/\xi$.}
\label{d1d2}
\end{figure}

The structure of the magnetic field of a hot star is typically not given by just
one multipole, but is much more complex
\citep[e.g.,][]{koc37776,marnyrus,silkor}. Therefore, we have to combine
multipoles with different orders to understand the formation of the equilibrium
surfaces in more realistic situations. In Fig.~\ref{mraknm} we provide plots of
individual equilibrium surfaces for different ratios of $B_m/B_n$ at the
Keplerian radius $\xi=1$ and for different orders $m$ and $n$ ($m>n$), assuming
that the multipoles share the axis of symmetry.

From Fig.~\ref{mraknm} it follows that the equilibrium surfaces are dominated by
lower-order multipoles for $B_m\lesssim B_n$. The presence of additional
multipoles is manifested by a slight shift of the equilibrium surface for
the case of multipoles with different parity. As the amplitude of the
higher-order multipole increases, a new equilibrium surface appears, which in the
limit $B_m\gg B_n$ approaches the structure of surfaces corresponding to the
order $m$ (cf. Fig.~\ref{mrakbrovno}). The new surface(s) has a toroidal shape
for multipoles of the same parity or a cone-like shape for multipoles with
different parity.

Figure~\ref{d1d2} explains why the lower multipole order dominates even for
$B_m\approx B_n$. Here we plot the magnetic strength vector for the dipolar and
quadrupolar component in the $y=0$ plane. The strength of the dipolar component is
always normalized to unity, while the quadrupolar component is plotted
relative to the dipolar component, that is, with the length reduced by a factor of
$1/\xi$. The total magnetic field strength is given by the sum of these
components. The figure shows that the quadrupolar component changes the orientation of the dominating dipole component only slightly in most cases.
This results in a warping of the equilibrium surface, but no new equilibrium
surface appears. Additional surfaces appear only when the quadrupolar
surface dominates the whole region.

These simulations show that a complexly warped light curve may appear only in
stars in which the higher-order multipole dominates even at the Keplerian radius
$\xi=1$. Because magnetic field components
corresponding to higher-order multipoles decrease fast, this implies an even stronger dominance at
the stellar radius. This condition can be partly mitigated by fast rotation, in
which case the stellar radius is close to the Keplerian radius.

\subsection{Nonaxisymmetric multipoles}
\label{kapneosym}

We have considered only axisymmetric multipoles so far. However, the general
multipole expansion also accounts for nonaxisymmetric terms
\citep[e.g.,][]{zahrada}. To understand their effect, we accounted for the
multipolar expansion in the form of 
\begin{multline}
\label{multipolna}
\boldsymbol{B}=
\frac{B_{n,l}}{r^{n+2}}P_n^l(\cos\theta)\cos(l\phi)\,\boldsymbol{e}_r-
\frac{1}{n+1}\frac{B_{n,l}}{r^{n+2}}\frac{\de P_n^{l}(\cos\theta)}{\de\theta}
\cos(l\phi)\, \boldsymbol{e}_\theta+\\
\frac{l}{n+1}\frac{B_{n,l}}{\sin\theta\,r^{n+2}}P_n^{l}(\cos\theta)\sin(l\phi)\,
\boldsymbol{e}_\phi.
\end{multline}

The nonaxisymmetric terms do not affect the shape of the equilibrium surface when the field is aligned with the rotational axis because the sum of the
gravitational and centrifugal force $\boldsymbol{f}$ has a zero azimuthal
component, and the radial and longitudinal field components both depend on $\phi$ in
the same way. As a result, the condition described by Eq.~\eqref{rovno} is not affected, and the
equilibrium surfaces are given by the radial and latitudinal components of the
magnetic field. The numerical analysis shows that the equilibrium surfaces
resemble those for multipolar fields given in Fig.~\ref{mrakbrovno}.

Because the individual magnetic field components depend on the radius in the same
way, the magnetic field lines are similar in the nonaxisymmetric case as well.
Therefore, the equilibrium surface takes a conical shape in the limit of large
$\xi$ for the same reasons as in the axisymmetric case. From a similar analytical
analysis as in Sect.~\ref{fiali} it follows that the radius of the central hole that
appears in the equatorial plane from stability considerations is not affected by
the azimuthal magnetic field and is given by Eq.~\eqref{blanskolibn}.

Because the number of lobes is given by the number of roots of the radial component
of the magnetic field, the number of equilibrium surfaces given by the numbers
of lobes could be higher than in the axisymmetric case. For instance, for $l=1,$
there is one additional root of the associated Legendre polynomial, therefore the
number of equilibrium surfaces is $n+1$. These surfaces show a mirror symmetry,
therefore for a field that is aligned with the rotational axis, the equatorial surface
fulfills the equilibrium condition for even $n$.

At the apex of individual lobes where $P_n^l(\cos\theta)=0$, only the
latitudinal magnetic field component remains. This component is modulated by
a factor of $\cos(l\phi)$, which has $2l$ roots. Therefore, there are $2l$
directions in each lobe where the flow can move freely and escape the star.
In reality, this picture may be modified by other field components. Because the
effective acceleration is not radial above the equatorial plane, the apex of
each lobe does not coincide with the equilibrium surface, as in the axisymmetric case.

For $\cos(l\phi)=0,$ only the azimuthal component of the magnetic field is
nonzero; consequently, $\boldsymbol{f}\cdot\boldsymbol{b}=0$ for aligned
rotational and magnetic axes, and new equilibrium surfaces appear. These surfaces
are planes with a common line of intersection corresponding to the
magnetic axis. The stability condition Eq.~\eqref{babice} for a particular case
of a nonaxisymmetric multipole as in Eq.~\eqref{multipolna} can after some
manipulation be rewritten as
\begin{multline}
\label{dlouhyvztah}
1-\frac{1}{\xi^3}-\zav{1-\frac{1}{\xi^3}-\cos^2\theta}\zav{n+2}+\\+
\cos\theta\zav{\frac{\sin\theta}{P_n^l(\cos\theta)}
\frac{\de P_n^{l}(\cos\theta)}{\de\theta}-\cos\theta}<0.
\end{multline}
For instance, in the equatorial plane $\cos\theta=0,$ this gives the condition $\xi>1$,
that is, the equilibrium is stable above the Keplerian radius. In the general
case, the plane is stable for large $\xi$. The numerical analysis has shown that
Eq.~\eqref{dlouhyvztah} allows a stable equilibrium even below the
Keplerian radius for $\xi<1$. With increasing complexity of the field (for
higher $n$), the smallest radius for which Eq.~\eqref{dlouhyvztah} is satisfied moves
toward the star, but it can never reach zero (as show below).

The stability condition has a particularly illuminating form for a purely
radial force (with $f_r(r)<0$) when the magnetic field is governed by a single
component. For radial force fields, the equilibrium condition Eq.~\eqref{rovno}
requires that the radial component of the magnetic field unit vector is zero,
$b_r=0$. This is fulfilled in the apex of magnetic lobes corresponding to the
roots of the Legendre polynomials $P_n^l(\cos\theta)$ (see
Eq.~\eqref{multipolna}). Because at the apex of the lobes, the longitudinal
magnetic field component is zero as well, $b_\phi=0$, the stability condition
Eq.~\eqref{stab} simplifies to
\begin{equation}
-b_\theta\frac{\partial b_r}{\partial\theta}<0.
\end{equation}
Inserting the multipolar field Eq.~\eqref{multipolna} and canceling the positive
terms, this can be rewritten as
\begin{equation}
\zav{\frac{\de P_n^l}{\de\theta}}^2<0,
\end{equation}
which is never fulfilled. Therefore, the stability in the case of a centered
magnetic field governed by a single component always requires the presence of
centrifugal force. A similar condition can be derived also for $\cos(l\phi)=0$
surfaces. For a general field, which can be described by a combination of
several components, the stability condition reads
\begin{equation}
b_\theta\frac{\partial b_r}{\partial \theta}+
\frac{b_\phi}{\sin\theta}\frac{\partial b_r}{\partial \phi}>0.
\end{equation}
Using the zero current condition $\text{rot}(B\boldsymbol{b})=0,$ this can be
rewritten as
\begin{equation}
b_\theta\frac{\partial}{\partial r}\zav{rBb_\theta}+
b_\phi\frac{\partial}{\partial r}\zav{rBb_\phi}
=\frac{\partial\zav{rB}}{\partial r}+\frac{rB}{2}\frac{\partial}{\partial
r}\zav{b_\theta^2+b_\phi^2}>0.
\end{equation}
This is difficult to fulfill for any radially decreasing magnetic field.
Therefore, either a centrifugal force or possibly nonzero currents are required
for stability.

\begin{figure}
\includegraphics[width=0.5\textwidth]{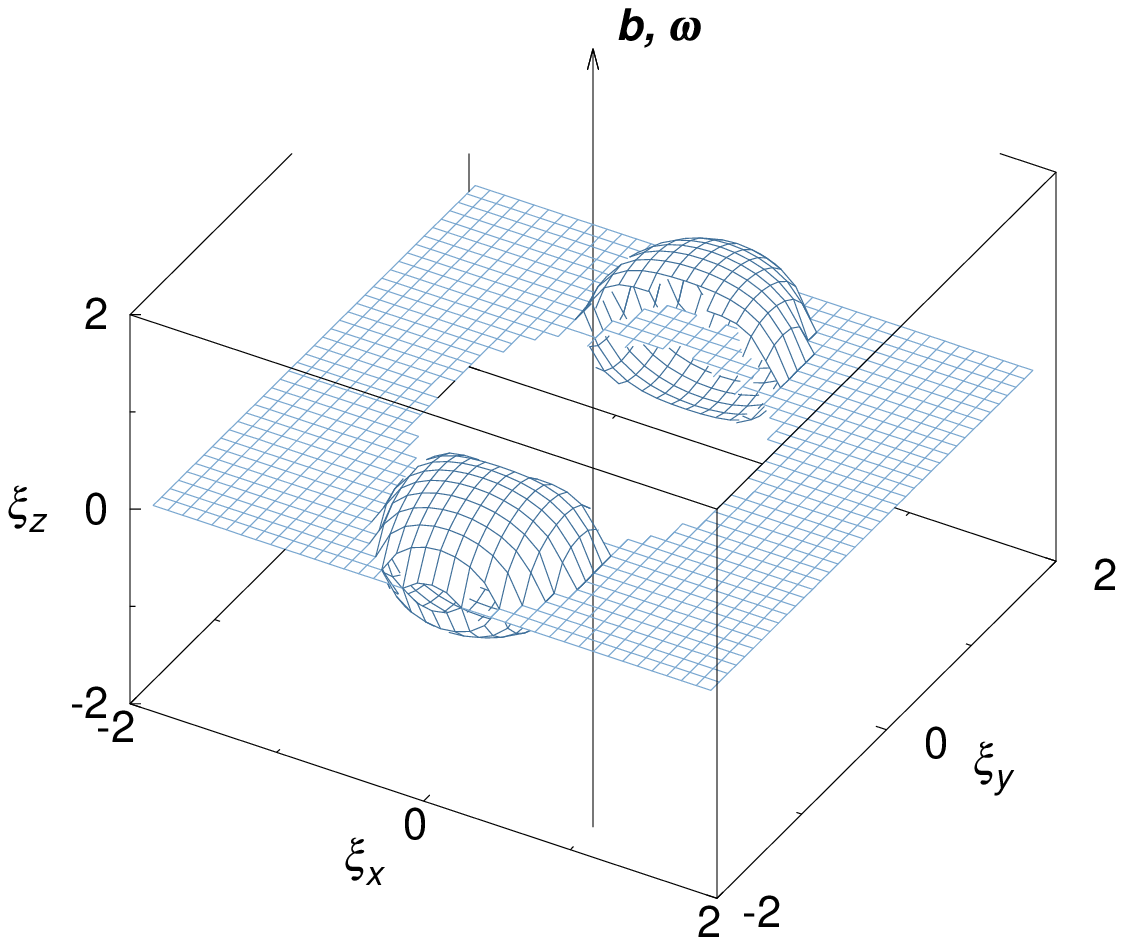}
\includegraphics[width=0.5\textwidth]{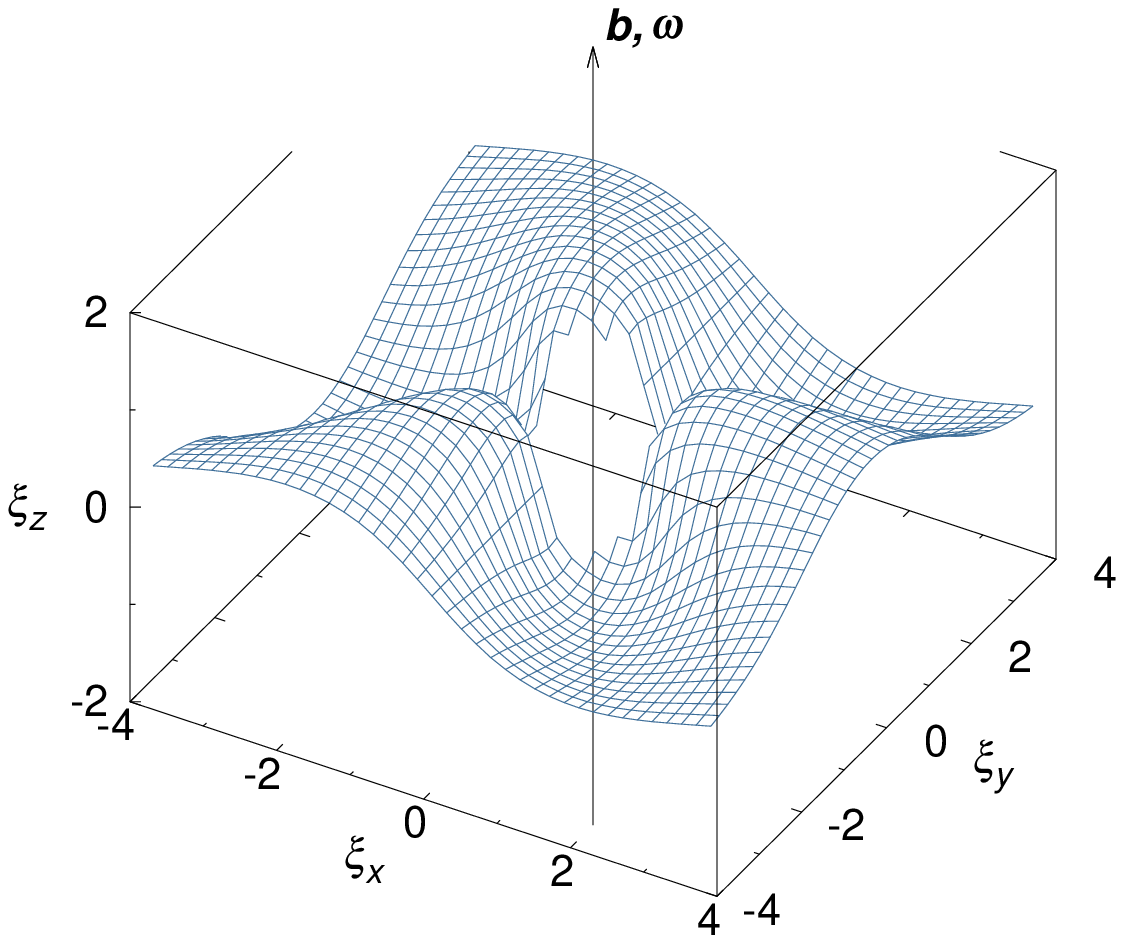}
\caption{Equilibrium surface fulfilling the stability condition for a field
aligned with the rotational axis and combination of the axisymmetric dipole and
octupole with $l=2$ ({\em upper panel}) and $l=3$
({\em lower panel}). Plotted for $B_{3,2}/B_1=0.3$ or $B_{3,3}/B_1=0.3$.}
\label{mraknan1m3}
\end{figure}

The field strength of the nonaxisymmetric multipoles varies with the same power
of radius as the strength of the axisymmetric multipoles. Therefore, low-order
multipoles dominate the magnetic field even in the nonaxisymmetric case unless
the expansion coefficients in Eq.~\eqref{multipolna} are sufficiently high.
However, the magnetic field component in the direction of force is zero around
the equilibrium surfaces. Consequently, even a small additional magnetic field
may dominate here and perturb the equilibrium surface. This is shown in
Fig.~\ref{mraknan1m3} (lower panel), where we plot the equilibrium surface for
the combination of the axisymmetric dipole and nonaxisymmetric octupole with
$l=3$. For a dipole, the equilibrium surface appears in the equatorial plane and
corresponds to the apex of the lobe of the magnetic field, where the radial field
component is zero. Consequently, even a relatively weak non-axisymmetric octupole
that lacks the apex of the lobe in the equatorial plane is able to warp the
equilibrium plane. Here the number of warps is equal to $l=3$.

Equilibrium surfaces become warped only if the surfaces of individual multipoles
do not coincide. In the opposite case, the splitting of surfaces may appear as a result
of the interaction of nearby equilibrium surfaces (see Fig.~\ref{mraknan1m3} for
$l=2$ case). Relatively complex shapes of the equilibrium surfaces can also be derived from a combination of multipoles with the same order, but with
different~$l$.

\subsection{Radius of the magnetosphere}

The magnetic field dominates the magnetosphere up to the Alfv\'en radius
$R_\text{A}$, at which the magnetic field energy density is equal to the stellar
wind kinetic energy density \citep{udo}. Therefore, the Alfv\'en radius can be
derived from the condition
\begin{equation}
\label{alfpod}
\frac{B^2}{8\pi}=\frac{1}{2}\,\rho\, v^2.
\end{equation}
For multipolar magnetic field $B\sim B_0(R_\ast/r)^{n+2}$ and stellar wind density
given by $\rho=\dot M/(4\pi r^2 v_\infty),$ at large distance from the star
($v=v_\infty$), the Alfv\'en radius is
\begin{equation}
\label{ralf}
\frac{R_\text{A}}{R_\ast}=\eta_\ast^{\frac{1}{2n+2}},
\end{equation}
where
\begin{equation}
\label{eta}
\eta_\ast=\frac{B_0^2R_\ast^2}{\dot Mv_\infty},
\end{equation}
which is the wind magnetic confinement parameter \citep{udo}. A strong
dependence of the Alfv\'en radius on the order of the multipole means that even in
the case of a very high magnetic field confinement $\eta_\ast\approx10^5-10^7$
\citep{biograf}, the Alfv\'en radius is just a few times higher than the stellar
radius for $n>3$. In this case, the Alfv\'en radius may become smaller than the
Kepler radius, preventing the existence of stable magnetospheric matter.

When the velocity is dominated by the rotational velocity, the condition
Eq.~\eqref{alfpod} should be modified to \citep{trimod} 
\begin{equation}
\frac{B^2}{8\pi}=\frac{1}{2}\,\rho\, v_\text{rot}^2(r).
\end{equation}
For solid-body rotation, $v_\text{rot}(r)=v_\text{rot}(R_\ast)r/R_\ast$, and
using the stellar wind density, the Alfv\'en radius is given by
\begin{equation}
\label{alfrot}
\frac{R_\text{A}}{R_\ast}=\eta_\ast^{\frac{1}{2n+4}}
\zav{\frac{v_\infty}{v_\text{rot}(R_\ast)}}^{\frac{1}{n+2}}.
\end{equation}
Because the ratio of the terminal wind speed and the equatorial rotational velocity typically is a factor of a few \citep{cak}, depending on the stellar parameters,
Eq.~\eqref{alfrot} could give an even smaller Alfv\'en radius than Eq.~\eqref{ralf}.
For higher-order multipoles, both expressions give results close to the stellar
radius.

Above the Alfv\'en radius, the magnetic field no longer dominates and the wind
flows nearly radially \citep{udo}. The region of equipartition between the magnetic
field energy density and the wind energy density around the Alfv\'en radius is the region of the reconnection events. They may accelerate particles to relativistic
energies \citep{trimod,letcuvir}.

\section{Light curves due to a rotating magnetosphere}
\label{krivulemulti}

The matter accumulates on equilibrium surfaces in the region of stability
\citep{towo}. The density is highest in the part of the surface that is
closest to the star. This material may obstruct the radiation emitted by the star
and obscure part of the stellar surface. Wind models \citep{metuje} show that
absorption is most likely dominated by the light scattering on free electrons.
The obscuration is highest for rays that are tangential to the
equilibrium surface, because these rays encounter the largest amount of the material.
As the star rotates, different parts of the equilibrium surface obstruct the
line of sight. Therefore, the obscuration depends on the rotational phase and
the light curve displays eclipses, as shown in \citet{towog}.

The analysis of high-precision satellite photometry revealed that some
chemically peculiar stars display small dips on their light curves.
These features look like absorption features and typically do not come in isolation, but
(when present) up to a dozen or more of such features appear in the light curve
\citep{mikland}. In our model, the number of these features per rotational
period increases with the complexity of the field, therefore with the order of the
multipole. From this, it seems that the warped light curves of chemically
peculiar stars can be explained by the light absorption in corotating clouds
that are trapped by the magnetic field that is described by multipoles of high order.

The magnetospheric matter settles in the magnetosphere, and its density
distribution is given by the hydrostatic equilibrium along each field line
\citep{towo},
\begin{equation}
\label{hustpsi}
\rho(\Delta s)=\rho_\text{m}\exp\hzav{-\frac{\mu\Delta\Phi(\Delta s)}{kT}},
\end{equation}
where $\Delta\Phi(\Delta s)$ is the difference between the potential at a given
point and the potential minimum located at the distance $\Delta s$ along the
field line, $\rho_\text{m}$ is the matter density at the potential minimum,
$\mu$ is the mean molecular weight, and $k$ is the Boltzmann constant.
Approximating $\Delta\Phi$ by its Taylor expansion and using Eq.~\eqref{stab},
$\Delta\Phi(\Delta s)\approx -f'\Delta s^2/2$. Therefore, the density
distribution Eq.~\eqref{hustpsi} takes the form of
\begin{equation}
\label{hustdisk}
\rho(\Delta s)=\rho_\text{m}\exp\zav{-\frac{\Delta s^2}{h^2}},
\end{equation}
with the square of the characteristic scale height,
\begin{equation}
\label{hand}
h^2=\frac{2kT}{\mu\azav{f'}}.
\end{equation}
The density $\rho_\text{m}$ is different for individual field lines, and
\citet{towo} accounted for stellar wind feeding to determine its value.
However, the observational characteristics of H$\alpha$ line profiles
\citep{shalfa} show that the magnetospheric matter density is given by a complex
interplay between wind feeding and gas leakage either via diffusion and drift
\citep{ocrali} or more likely by centrifugal breakout \citep{ohalfa}.
Consequently, we simply assumed $\rho_\text{m}=A\xi^{-6}$ \citep{ohalfa}, where
$A$ determines the total amount of mass in the magnetosphere.

\subsection{Light absorption due to the circumstellar magnetosphere}

As the circumstellar matter obstructs the rays aimed at a distant observer, it
absorbs the stellar radiation and causes the light variability. The intensity
along the ray is reduced according to
\begin{equation}
\label{itau}
I=I_0 e^{-\tau},
\end{equation}
where the optical depth is given by
\begin{equation}
\label{tau}
\tau=\int \frac{\sigma_\text{T}\rho}{m_\text{H}}\,\de l=
\frac{\sigma_\text{T}A}{m_\text{H}}
\int \frac{1}{\xi^{6}} \exp\zav{-\frac{\Delta s(l)^2}{h^2}}\de l,
\end{equation}
where we used Eq.~\eqref{hustdisk}, $l$ denotes the length variable along the ray,
$m_\text{H}$ is the hydrogen mass, and $\sigma_\text{T}$ is the Thompson
scattering cross-section. We integrated the specific intensity from
Eq.~\eqref{itau} across the visible stellar surface for different viewing angles
to obtain the phase-dependent light curve.

\begin{figure}
\includegraphics[width=0.5\textwidth]{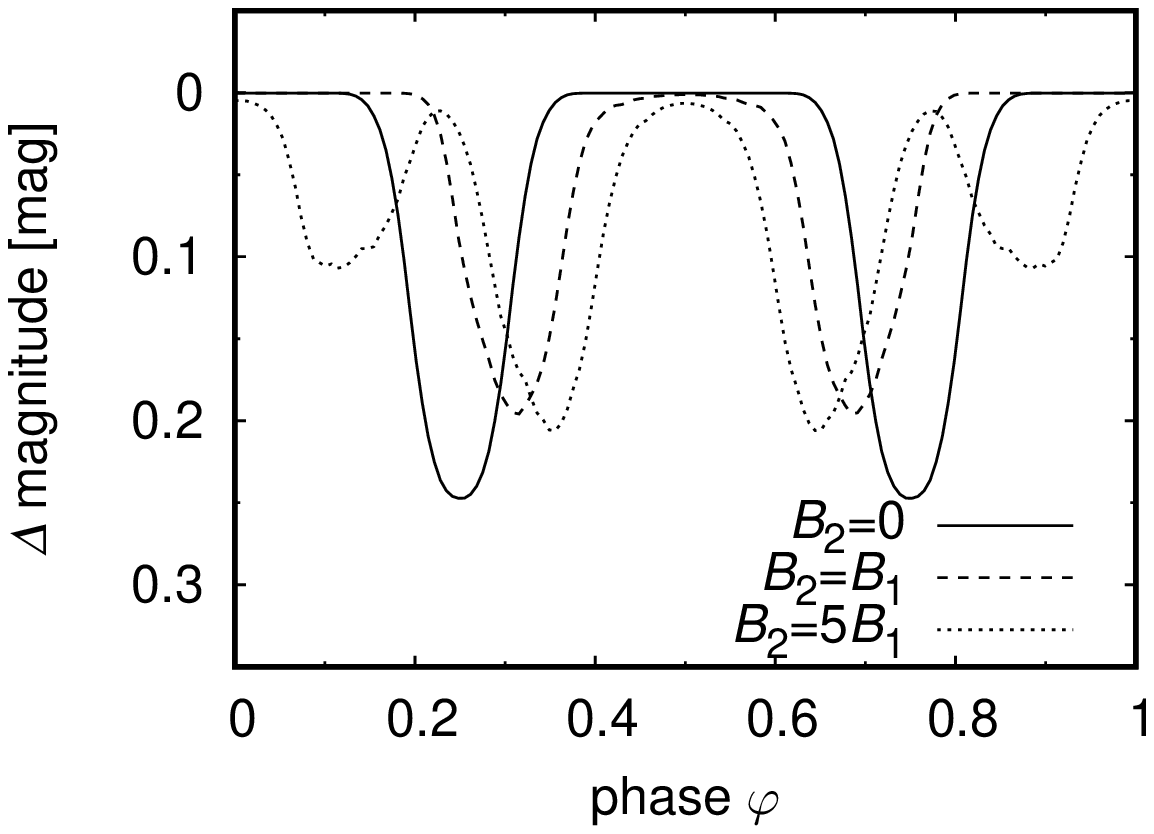}
\includegraphics[width=0.5\textwidth]{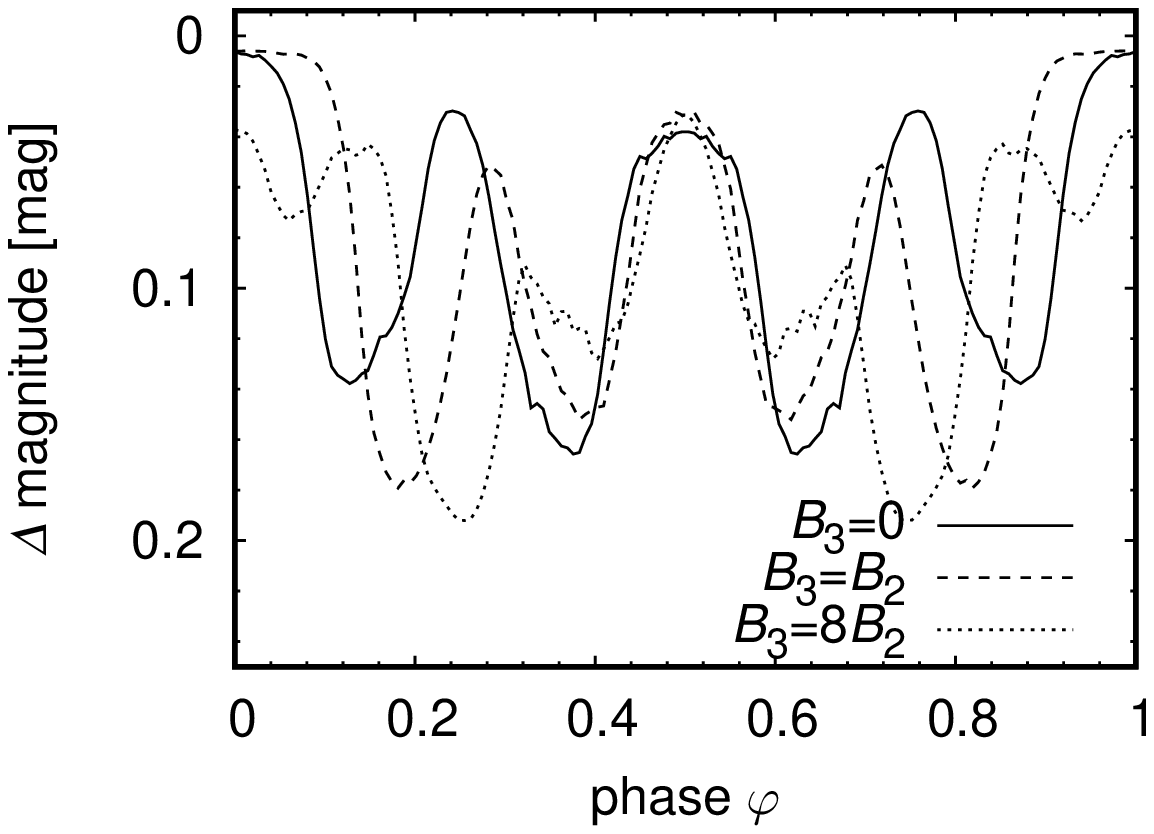}
\includegraphics[width=0.5\textwidth]{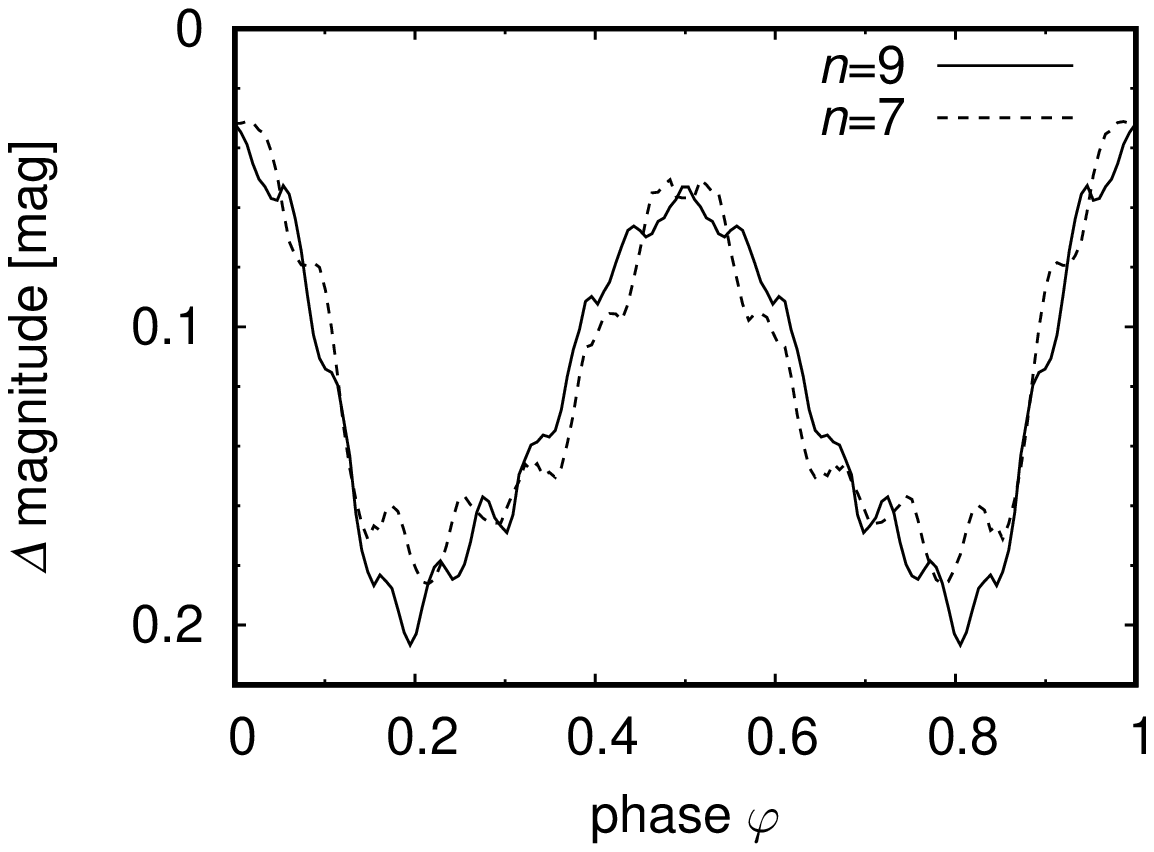}
\caption{Light curves due to the light absorption in circumstellar
magnetospheres for $i=\beta=\pi/2$. {\em Upper panel}: Combination of dipole
and quadrupole. {\em Middle panel}: Combination of quadrupole and octupole.
{\em Bottom panel}: Light curves due to higher-order multipoles.}
\label{krivky}
\end{figure}

The magnetospheric matter accumulates around the equilibrium surfaces
Eq.~\eqref{rovno}. For an axisymmetric magnetic field with $\beta=0,$ the resulting
density distribution is also axisymmetric. Consequently, there is no rotational
variability in the case of aligned rotation of axisymmetric multipoles. For
$\beta>0,$ the equilibrium surfaces become warped even for the case of multipoles
that are axially symmetric around the magnetic axis. Most of the matter accumulates in
the surface regions closest to the star, leading to a modulation of the light variability by the rotation.

Figure~\ref{krivky} shows light curves for different combinations of multipoles.
We adopted a stellar mass $M=8\,M_\odot$, a radius $R=4\,R_\odot$, and a rotation
period $P=1\,\text{d}$ roughly corresponding to $\sigma$~Ori~E
\citep{mysigorie}. The light curves are plotted for a magnetic axis perpendicular
to the rotational axis ($\beta=\pi/2$) and for an equator-on orientation
($i=\pi/2$). This gives the largest amplitude of the variability \citep{towsam}.
With a nonzero magnetic axis tilt, the equilibrium surfaces become warped with
the maximum density of matter in the surface regions closest to the star. As the
equilibrium surfaces show a mirror symmetry around the plane containing the rotational
and magnetic axes (see Fig.~\ref{mrakpsi75}), there are two magnetospheric
clouds per equilibrium surface. Therefore, each equilibrium surface produces two light minima at most, which appear when the cloud occults part of the stellar
surface. Because the number of surfaces is given by the order of the multipole,
the maximum number of light curve minima is twice the order of the multipole.
However, minima due to higher multipoles show up only when these multipoles
dominate at the Kepler radius. When the higher-order multipoles do not dominate
at the Kepler radius, then they cause just a phase shift of minima due to
equilibrium surface elevation (see Fig.~\ref{mraknm}).

For higher-order multipoles ($n>3$), the rays that are neither close to
parallel nor normal to the rotational axis may intersect several equilibrium
surfaces. Consequently, the minima due to individual surfaces merge and create
one huge absorption feature that is modulated by the minima due to individual
surfaces (Fig.~\ref{krivky}).

We also tested the influence of nonaxisymmetric multipoles
(Sect.~\ref{kapneosym}) on the light curve. Inclusion of the azimuthal magnetic
field may increase the number of equilibrium surfaces and therefore the number
of warps. Higher-order multipoles are still needed to obtain a large number
of warps in the light curve, however. The equilibrium surfaces are axisymmetric
even for nonaxisymmetric multipoles for a field that is aligned with the
rotational axis ($\beta=0$) because the alignment forms a disk-like
circumstellar structure. However, because the $f'$ term in the disk density
scale height Eq.~\eqref{hand} depends on mixed terms, the density distribution
is not axisymmetric for equilibrium surfaces that do not lie in the equatorial
plane. As a result, the variability appears even in the case of a field that is
aligned with the rotational axis.

The disk width is inversely proportional to the square root of $|f'|$.
Therefore, the disk width is large (disk flares) when $f'$ tends to zero. This
leads to an occultation of a large part of the stellar surface and to the appearance of
a minimum in the light curve. For nonaxisymmetric multipoles, this appears when
$\cos(l\phi)=0$. Therefore, there are $2l$ such flaring regions that may lead to
$2l$ absorption features in the light curve.

\begin{figure}
\includegraphics[width=0.5\textwidth]{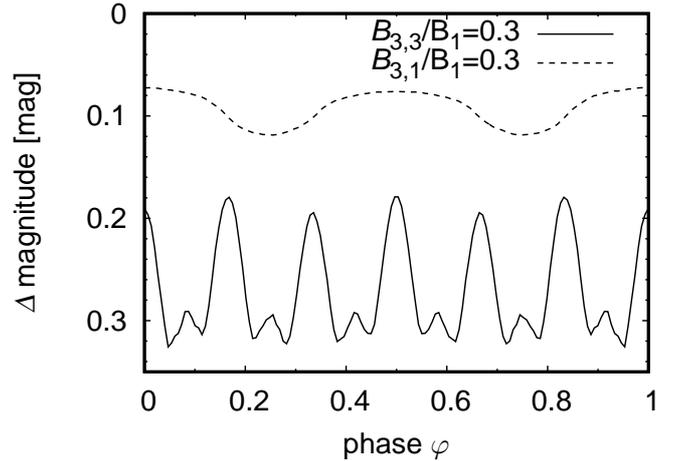}
\caption{Light curves due to the light absorption in circumstellar
magnetospheres for a combination of the axisymmetric dipole and octupole with $l=3$
and $l=1$ for $\beta=0$ viewed equator-on ($i=\pi/2$).}
\label{krivkyna}
\end{figure}

Even a tiny nonaxisymmetric component in an otherwise dominant
axisymmetric field leads to a warping of the equilibrium surface even
when the rotational and magnetic axes coincide (Fig.~\ref{mraknan1m3}). This
results in a light curve with $2l$ dips caused by $2l$ disk warps
(Fig.~\ref{krivkyna}).

\subsection{Continuum emission from a corotating magnetosphere}

Part of the light absorbed by the corotating magnetosphere is emitted again and
may reach a distant observer \citep{mysigorie}. If the extinction appears due
to light that scatters on free electrons, then the scattered light has the same
spectral energy distribution as the star and there is no net absorption in the
magnetosphere. However, unlike the simulation of the magnetospheric light
absorption, modeling the light emission is a formidable problem. With
emission, the radiative transfer equation needs to be solved in its full
integro-differential form for all rays intercepting all magnetospheric points.
In addition, the solution of the radiative transfer equation should be
iterated for a consistent solution.

To make the problem more tractable, we did not solve the radiative transfer
equation within the clouds and simply assumed that the radiation is reflected by
the surface of the clouds. The amount of energy emitted by the stellar surface
element $\de S\!_\ast$ is $F_0\,\de S\!_\ast$, where $F_0$ is the surface flux.
Assuming that the stellar surface radiates according to the cosine law, the flux
observed at the the distance $r_\ast$ from the surface element is $F_0\,\de
S\!_\ast \cos\theta_\ast/(\pi r_\ast^2)$, where $\theta_\ast$ is the angle
between the normal to the surface and the ray. A geometrically thin magnetospheric
medium can be described by its cross-section with respect to the infalling
radiation $\cos\upsilon\,\de S$, where $\upsilon<\pi/2$ is an angle between
the normal to the equilibrium surface element (given by
$\nabla(\boldsymbol{f}\cdot\boldsymbol{b})$ from Eq.~\eqref{rovno}) and the
direction to the stellar surface element. The amount of radiation that is scattered by a
given element is then $\cos\upsilon\,\de S(1-e^{-\tau})F_0\,\de
S\!_\ast\cos\theta_\ast/(\pi r_\ast^2)$. Assuming that the light is
redistributed in the magnetospheric matter according to the cosine law, the
flux observed at the distance $d$ from the star is $\cos\theta\cos\upsilon\,\de
S(1-e^{-\tau})F_0\, \de S\!_\ast\cos\theta_\ast/(\pi^2 r_\ast^2d^2)$, where
$\theta<\pi/2$ is the angle between the normal to the surface and the direction to
the observer. Integrating over all stellar and equilibrium surfaces, the total
observed reemitted flux relative to the flux coming from the star $F=
R_\ast^2/d^2 F_0$ is
\begin{equation}
\label{emvzor}
\frac{\Delta F}{F}=\int\de S \cos\theta\int (1-e^{-\tau})
\frac{\cos\upsilon\cos\theta_\ast}{\pi^2 r_\ast^2R_\ast^2}\,\de S\!_\ast.
\end{equation}
When we numerically evaluated Eq.~\eqref{emvzor}, we assumed that each surface
element directly faces the star, and we accounted for occultation by the star. To
simplify the calculation, the optical depth was approximated from Eq.~\eqref{tau}
by
\begin{equation}
\tau=
\frac{\sigma_\text{T}A}{m_\text{H}}
\int \frac{1}{\xi^{6}} \exp\zav{-\frac{l^2}{h^2}\cos^2\upsilon}\de l\approx
\frac{\sqrt{\pi}\sigma_\text{T}Ah}{m_\text{H}\xi^{6}\cos\upsilon}.
\end{equation}

The test showed that the total emitted flux calculated with this method is lower than the total absorbed flux by just a
few percent. This is an acceptable difference
because absorption and emission were treated differently, which is not fully
compatible.

\begin{figure}
\includegraphics[width=0.5\textwidth]{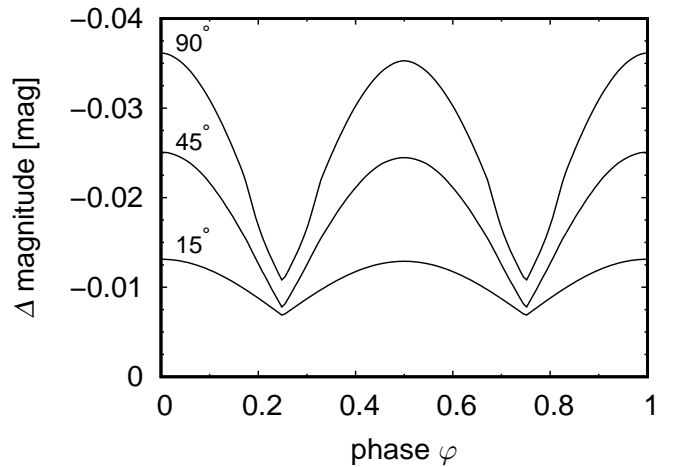}
\caption{Light curves due to the light emission in circumstellar
magnetospheres for a dipolar magnetic field with $\beta=90^\circ$ and different
inclinations denoted in the graph.}
\label{krivkyem}
\end{figure}

Figure~\ref{krivkyem} shows light curves due to magnetospheric emission for the
case of a dipole ($n=1$) with a rotational axis perpendicular to the magnetic
field axis ($\beta=90^\circ$) and for different inclinations. The light curve
due to light emission has a smooth profile and the variations appear because the
line of sight changes. The minimum has a wedge-like shape with a nearly linear
behavior on either side of the minima because the $\cos\theta$ term appears in
Eq.~\eqref{emvzor}, which is roughly proportional to $|\cos\varphi\,|$. The
maxima due to each cloud are shifted with respect to the absorption minima by
$0.25$ in phase because the normal to the equilibrium surface is nearly
perpendicular to the rotational axis and the maximum appears at the phase when
the surface normal points in the direction of the observer. In contrast to the
absorption, the variability appears even for a small magnetic field tilt and for
a small inclination \citep[cf.][]{towsam}.

%

\section{Application to stars with warped light curves}
\label{dalsi}

\begin{figure}
\includegraphics[width=0.5\textwidth]{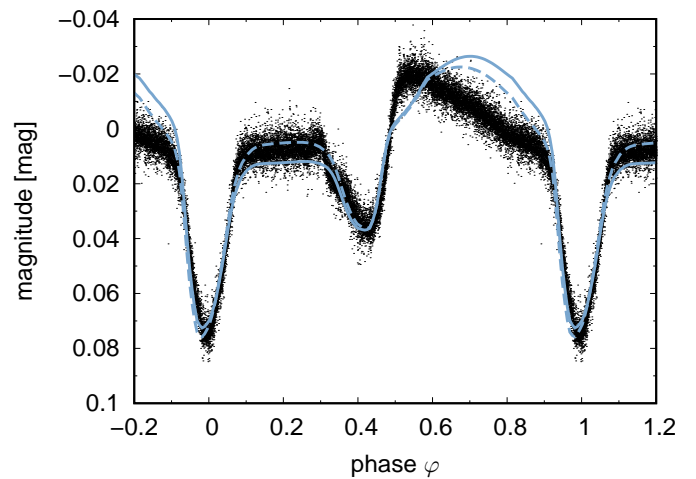}
\caption{Simulated light curve assuming a magnetic field given by the combination of
a dipole and quadrupole and accounting for both absorption and emission in the
magnetosphere (dashed line). The solid line corresponds to the model that also
accounts for surface spots \citep{mysigorie}. This is compared with the observed light curve
of $\sigma$ Ori E \citep{promneprom} and phased with the ephemeris from
\citet{town}.}
\label{sigorie}
\end{figure}

\subsection{Star $\sigma$ Ori E}

The enigmatic light curve of \object{$\sigma$ Ori E} \citep[see also
Fig.~\ref{sigorie}]{hesbin} motivated several studies that interpreted its light
variability as the result of absorption in the circumstellar magnetosphere
\citep{labor,nakaji,prus,towog}. The light curve shows two deep minima that
originate from the absorption in the stellar magnetosphere and from additional
variability between the minima, possibly due to the light scattering
\citep{mysigorie}. With respect to the light curves of other chemically peculiar
stars, it might be regarded as a warped light curve at its extrema.

We modeled the light curve assuming an inclination $i=75^\circ$, which roughly
corresponds to observations \citep{mysigorie}, and a magnetic field tilt
$\beta=90^\circ$. We selected a higher value of the tilt than derived from magnetic
maps \citep{mysigorie} to obtain a light curve maximum height that agrees with
observations. As a result of the magnetic field tilt, the circumstellar matter
is not distributed axisymmetrically, but mostly appears at the intersection
of the rotational and magnetic equatorial surfaces \citep{towog}. This results in
two circumstellar clouds that periodically occult the stellar surface, leading
to the light curves with two minima. To fit the width of the minima, we assumed
the magnetospheric density parameter $\rho_\text{m}=A\xi^{-3}$. This corresponds
to the stellar wind attenuation in a dipolar field and gives a slower decrease in
density than predicted by the model of a centrifugal breakout \citep{ohalfa} that we used
in our previous calculations. This particular choice of the radial power-law
index is also motivated by observational data. A higher index of the radial power law
predicts a wider occultation phase than observed because most of the absorption
appears closer to the star. The observed minima do not appear shifted by 0.5 in
phase, which indicates that the magnetospheric matter is not distributed
symmetrically \citep{towog}. Therefore, we assumed a combination of the dipolar and
quadrupolar components $B_2/B_1=0.6$ to shift the equilibrium surface (see
Fig.~\ref{mraknm}) and consequently also the light minima in accordance with
observations. A similar approach was used by \citet{mysigorie}. Additionally, we
introduced a dependence of the density parameter $A$ on a position to account for
different depths of light minima. We assumed that the value of $A$ in one
half-space of the magnetosphere is half that in the second half-space.

The simulated light curve in Fig.~\ref{sigorie} reproduces the main features of the
observed light curve, although some details remain unexplained. The deep minima
are nicely reproduced. The variations between the minima can be interpreted as
due to the light scattering in the magnetosphere. As a result of the shift in
the equilibrium surface introduced above, only one side of the surface is
illuminated by the star, therefore the light curve due to the scattering
contains only one maximum. This broad maximum appears around the phase
$\varphi=0.7$ and is the only effect in the light curve due to the light
scattering. From this it follows that the quadrupolar component is needed not only
to shift the light minima in phase, but also to explain the different heights of the
local maxima in the light curves. Compared to the circumstellar density
distribution determined by \citet{mysigorie}, the density distribution is
asymmetric because of the variation in $A$ parameter that we adopted, and it is more aligned with
the rotational axis.

The shift between the predicted and observed maxima of
about 0.2 in phase (Fig.~\ref{sigorie}) remains to be explained. It likely appears because the shape of the equilibrium surface is more complex than expected. We have not been able to
explain this shift by variations in any magnetospheric parameters, including
different orders of multipole, their mutual strength, and magnetic field
inclination. The shift between primary minima and maxima due to emission is
nearly 0.5 in phase, which means that it can be explained by assuming that the
surface around which the magnetospheric cloud accumulates is perpendicular to
the radial direction (i.e., the surface forms a small section of a sphere). A
similar model was introduced to explain the continuum polarization in this star
\citep{carsigopol}. The adopted model of geometrically thin clouds
perhaps oversimplifies the situation.

The model of two clouds connected by a ring proposed by \citet{carsigopol} might be interpreted assuming a nonaxisymmetric field with $n=l=1$. This model
leads to the appearance of two flaring regions, which would explain the two deep minima.
The flaring regions are extended in the direction perpendicular to the radial
direction, therefore the extrema due to light emission and absorption coincide
in phase.

\citet{mysigorie} introduced another component of the light variability of
$\sigma$ Ori E due to abundance spots that appear on the surface of this star
and that can be revealed from Doppler mapping. The abundance spots dominate the
light variability in the ultraviolet domain to a great extent, and they manifest
themselves as an additional shallow light maximum around phase 0.9 in the
optical region. Accounting for the spots in our model has a small impact in the
light curve, except for a slight shift of the light maximum toward a higher
phase (Fig.~\ref{sigorie}).

We aimed to determine a magnetic field distribution that was able to reproduce
not only eclipses, but also the height of the emission-like feature around phase
$\varphi=0.6$. In turn, the resulting surface magnetic field poorly
agrees with the observed longitudinal magnetic field curve. This again indicates that further improvement of the magnetospheric model is required.

\subsection{HD 37776 and other stars with warped light curves}

\begin{figure}
\includegraphics[width=0.5\textwidth]{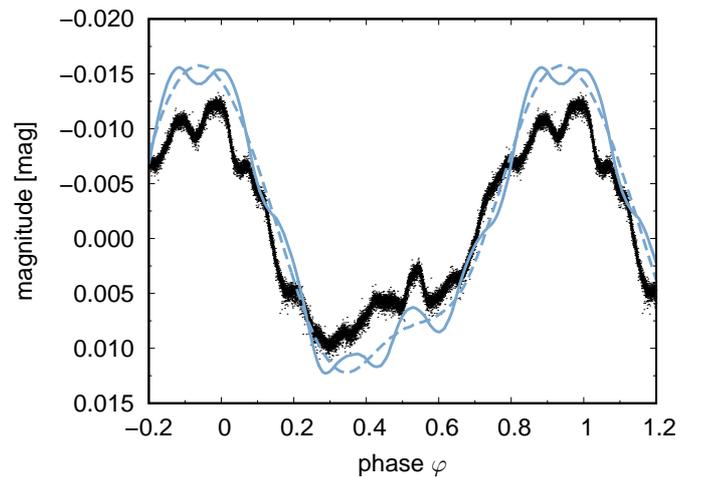}
\caption{Simulated light curve of HD~37776 derived assuming the silicon and helium
surface distribution by \citet{choch} and modulated by $n=3$ and $l=3$
multipole (solid line). This is compared with the TESS light curve phased with
the nonlinear ephemeris from \citet[dots]{mikland}
and with the light curve calculated purely from surface spots (dashed line). All
light curves were shifted to derive a zero mean magnitude.}
\label{hd37776}
\end{figure}

The light curve of \object{HD 37776} can mostly be explained as a result of flux
redistribution in surface abundance patches of helium and silicon
\citep{myhd37776}. A detailed inspection of the HD~37776 light curve derived
with the TESS satellite, however, revealed about ten localized narrow features
that can be interpreted in the form of dips
\citep[warps;][Fig.~\ref{hd37776}]{mikland} with additional substructures.
\citet{mikland} succeeded in providing a detailed phenomenological model of the
HD~37776 TESS light curve, but the physical origin of the fine structure
remained unclear. The star shows a complex magnetic field \citep{thola} with a
dominating octupolar component \citep{koc37776} and a significant nonpotential
field ($\text{rot}\,\boldsymbol{B}\neq0$). Some of the strong warps might be
attributed to the dominating $l=3$ nonaxisymmetric component of the field. This
component should give rise to six warps, which is slightly fewer warps than
derived from observations.

This is illustrated in Fig.~\ref{hd37776}, where we plot HD~37776 light
curve\footnote{The detailed ephemeris will be published elsewhere (Mikul\'a\v
sek et al., in preparation). The phase of the adopted nonlinear ephemeris can be
approximated as
$\vartheta=\vartheta_0-\dot{P}_0\,\vartheta_0^2/2-P_0\ddot{P}_0\,\vartheta_
0^3/6$, where $\vartheta_0 = (t-M_0)/P_0$. $P_0 = 1.538\,736(2)$ days,
$\dot{P}_0 = -1.51(4) \times 10^{-8}$, $\ddot{P}_0=-3.07(12)\times
10^{-12}$\,d$^{-1}$, and $M_0 = 2\,459\,580.715(5)$.} simulated by
\citet{myhd37776} from the \citet{choch} abundance maps, which was further modulated
by circumstellar absorption due to $n=3$ and $l=3$ aligned multipoles
($\beta=0$). In Sect.~\ref{kapneosym} we identified additional equilibrium
planes $\cos(l\phi)=0$  that intersect the magnetic field axis. These surfaces
appear in the region in which the equilibrium disk flares, therefore we did not
account for the variability induced by them. The comparison with the light curve
derived using the TESS satellite \citep{commander} shows slightly higher amplitude
of the simulated data. This might be attributed to a missing iron opacity,
which was not accounted for in our atmosphere modeling that we used to determine
emergent fluxes, and it can compete with other opacity sources. Inclusion of
circumstellar absorption better reproduces the overall shape of the light curve,
although many details are missing, which indicates that shape of the
magnetosphere might be more complex. This might be connected with additional components of
the magnetic field. Alternatively, the light curve could be interpreted as a result
of light absorption on the equilibrium surface of low-order axisymmetric multipoles
(dipole or quadrupole) warped by a higher-order nonaxisymmetric multipole. We
also note that the adopted magnetic field model was not selected to reproduce the
surface magnetic field distribution derived from spectropolarimetry, and a
more realistic magnetic field model may fit the observed light curve better.

The magnetic fields of most magnetic, chemically peculiar stars are dominated by
the dipolar component already at the stellar surface
\citep{spotil,dvojka,marnyrus,shumagrot,kotep}, and stars in which higher-order
multipoles prevail are scarce \citep{thola,dontausco,bagr,jaknedip}. In stars
with purely dipolar components, simple light curves with at most two
absorption dips are predicted, such as that found in \object{$\sigma$ Ori E}
\citep{towog}. More complex light curves can appear in stars in which the higher-order multipoles dominate not only at the stellar surface, but at the Keplerian
radius, where the clouds start to form. This is an even more strict condition than
a dominance at the stellar surface caused by the radial decrease of higher-order
multipoles, which is faster than for a pure dipole.

It follows from this that the multiple warps in the light curves of stars with
prevailing dipolar magnetic field need to be explained by something else than the axisymmetric
model of a rigidly rotating dipolar magnetosphere. For instance, the perturbation of
an otherwise dominant axisymmetric field by even a weak nonaxisymmetric component
leads to the appearance of warps on the equilibrium surface (Fig.~\ref{mraknan1m3})
that may cause dips in the light curve (Fig.~\ref{krivkyna}). Given the
dominance of dipolar fields among magnetic, chemically peculiar stars, this
appears to be the most promising model of the nature of the tiny features in the
light curves of chemically peculiar stars.

On the other hand, if the magnetic field in the outer parts of the magnetosphere
has a more complex topology than a simple dipole and is, for instance, governed
by higher-order multipoles or has a component with a significant nonaxisymmetric
term or with nonzero rotation, then this field may lead to a more complex
distribution of magnetospheric matter and to the appearance of warps in the light
curve. In the context of stellar interiors, \citet{brano} showed with their
numerical simulations that stable internal fields are not composed just from
axisymmetric dipole, but also include a significant toroidal 
component. Magnetic field components with nonzero rotation may appear, for
example, due to outflows or due to electric currents flowing in the
magnetosphere. In this case, the explanation of warps by corotating
magnetospheric clouds might be possible even for stars with simple surface
fields. We additionally tested this possibility and simulated the light curve
due to a magnetosphere governed by higher-order multipoles with accumulating
clouds located in the outer regions of magnetosphere. The derived light curves
resembled the warped light curves, giving this possibility some credit.

Magnetic Doppler imaging has unveiled a magnetic field with nonzero
rotation in some hot stars \citep{koba53cam,dontausco,koc37776}, which might indicate nonzero electric currents. We tested the possibility that these
currents may be generated by nonzero velocity differences between oppositely
charged particles in line-driven wind. We used our multicomponent wind models
\citep{kkii} to estimate the magnitude of the electric current in B star winds. Our
models showed that the magnetic fields induced by wind currents are weaker by several
orders of magnitude than the field observed in magnetic, hot stars. Therefore,
it is more likely that the magnetic fields with nonzero rotation are connected
with atmospheric currents \citep{wolt,rakosnik} that are thought to contribute
to hydrogen and helium line profile variations \citep{madbal,valmaghyd,valli}.

If the magnetospheric clouds appear at larger distances where the magnetic field no longer corresponds to a dipole, then the centrifugal acceleration
dominates gravity in these regions. The vertical scale height at large
distances is given by
\begin{equation}
H=\frac{kT}{\mu f}=\frac{kT}{\mu\Omega^2 r}.
\end{equation}
In other words, as a result of the linear increase in centrifugal acceleration with
radius, the vertical scale height is inversely proportional to radius.
Therefore, we expect larger clouds close to the star and smaller ones at large
radii.

\section{Discussion: Other possible sources of complex light variations}

Matter cannot be supported by continuum radiation. This can be seen from
evaluating the maximum mass that can be supported radiatively, which is given by
the balance of the gravitational and radiative force ${GM\Sigma}/{R_\ast^2}={\sigma
T_\text{eff}^4}/{c}$, where $\Sigma$ is the column mass. The optical depth of
this material is given by ${\sigma_\text{T}\Sigma}/m_\text{H}=
{\sigma_\text{T}L}/\zav{4\pi c m_\text{H}GM}=\Gamma$. Therefore, the optical
depth is given by the Eddington parameter, which is $\Gamma<1$ for normal stars.
Consequently, a star cannot radiatively support material that is optically thick
due to Thomson scattering.

Some stars show a complex magnetic field structure just above the stellar surface,
with magnetic field lines closing significantly below the Keplerian corotation
radius \citep[e.g.,][]{kowy}. These structures do not necessarily lead to
corotating clouds supported by centrifugal force, as studied here. Instead, the
field lines may be filled with wind material, which collides with streams from
the opposite footpoint of the magnetic loop \citep{udorot,malykor}. This leads to
the appearance of a dynamical magnetosphere with a complex upflow and downflow
structure. These structures are inevitably unstable. However, because the flow
along different loops is independent, the combination of absorption from a large
number of loops may lead to a more or less time-independent photometric signature
even for a dynamical magnetosphere. Moreover, this effect might be stronger in cooler stars with a lower wind density because the
cooling time in these stars is longer and the post-shock gas is able to occult a larger fraction
of the stellar surface.

The density of the post-shock in nearly isobaric matter can be estimated as
\citep{andyma}
\begin{equation}
\rho_\text{h}=4\rho_\text{w}\frac{T_\text{s}}{T_\text{h}},
\end{equation}
where $\rho_\text{w}$ is the wind density, $T_\text{s}$ is the shock
temperature, and $T_\text{h}$ is the temperature of the cooled post-shock gas. The
Thomson scattering optical depth due to this structures is
\begin{multline}
\tau_\text{h}\approx \frac{\sigma_\text{T}}{m_\text{H}}
\frac{\dot M}{\pi v_\text{w} R_\ast^2}
\frac{T_\text{s}}{T_\text{h}}R_\text{h}=\\10^{-3}
\zav{\frac{\dot M}{10^{-10}\,M_\odot\,\text{yr}^{-1}}}
\zav{\frac{v_\text{w}}{100\,\kms}}^{-1}
\zav{\frac{R_\ast}{R_\odot}}^{-2}
\zav{\frac{R_\text{h}}{R_\odot}}
\frac{T_\text{s}}{T_\text{h}},
\end{multline}
where $R_\text{h}$ is the length corresponding to the post-shock gas, and
$v_\text{w}$ is the wind velocity. Because ${T_\text{s}}/{T_\text{h}}$ may be up
to $10^3$, the dynamical magnetosphere may affect the light curve only for stars
with a higher mass-loss rate $\dot M\gtrsim10^{-10}\,M_\odot\,\text{yr}^{-1}$.
This overcomes the mass-loss rates of a typical chemically
peculiar star by several orders of magnitude \citep{metuje}.

It is also possible consider the problem of the shock-heated magnetospheres
\citep{bamo,udomax} from the point of view of X-ray energetics. The density
scale height at these temperatures is comparable to the stellar radius,
therefore this matter is close to the hydrostatic equilibrium along the field
lines \citep{andyma}. The denser structures in this environment may give rise to the
warped light curves. The hot magnetospheric matter needs to obscure about
$\delta=10^{-3}$ of stellar radiation to cause warp features with an amplitude
of about 1\,mmag. This gives the condition for the minimum density of this
material as ${\sigma_\text{T}\rho}C/m_\text{H} \approx\delta$, where $C$ denotes
the length of the cloud in the direction of observation. This implies that the mass
of the cloud $ABm_\text{H}\delta/\sigma_\text{T}$ must be about
$10^{-14}\,M_\odot$ assuming the cloud dimensions $A\approx B\approx R_\ast$ and
the total mass of this corona-like structure of $10^{-13}\,M_\odot$ with about
ten such clouds. The X-ray luminosity of such a cloud would be about
$L_\text{X}\approx ABC \Lambda (\rho/m_\text{H})^2$, which, using the unity
optical depth condition and assuming $B\approx C,$ gives an X-ray luminosity of
$L_\text{X}\approx A\delta^2\Lambda/\sigma_\text{T}^2$. The X-ray cooling
function $\Lambda$ from \cite{jiste} gives
$L_\text{X}\approx10^{31}\,\text{erg}\,\text{s}^{-1}$, which is higher by two to three
orders of magnitude than the typical luminosity of magnetic B stars
\citep{nazmagx} and cannot be sustained by the weak winds
of late-B stars.

Prominence-like structures \citep[e.g.,][]{sudhen} may provide an analogy for the
proposed model of matter distribution leading to dips in the light curves.
Classical solar prominence models \citep{einekis,kuperuraad} nicely correspond
to structures that may cause warped light curves. The potential extrapolation of
surface fields in cool stars showed that equilibrium regions may exist even
below the Keplerian radius \citep{zahradajakomy} if the field is
sufficiently complex. However, the height of these structures would have to be
comparable to the stellar radius to cause warp features, which again requires higher-order multipoles.

\section{Conclusions} \label{zaver}

We simulated light curves due to light absorption and emission in the matter
that is centrifugally supported in magnetospheres for a magnetic field governed by higher
orders of multipolar expansion. Our aim was to understand the complex light
curves of chemically peculiar stars, which show persistent phased multiple
features that can be interpreted as dips (warps).

We have shown that with increasing order of the multipolar expansion of the
magnetic field, the complexity of the light curve increases. Two warps
at most appear per order of multipolar expansion for axisymmetric fields. However,
higher-order axisymmetric multipoles have to dominate not only at the stellar
radius, but also at the Keplerian radius to significantly affect the light
curve. A similar condition was found by \citet{zahradajakomy} for the existence of
prominences in cool stars. Because most of the hot, magnetic stars do not show these intricate surface fields, complex warped light curves originating from
complicated surface fields are expected to be rare. A study is under way to test
this prediction (Mikul\'a\v sek et al., in preparation).

We distinguished two geometrically different sources of variability. For axially
symmetric magnetic fields, the resulting distribution of magnetospheric matter
retains some kind of symmetry, therefore a nonzero tilt between the magnetic and
rotational axes is needed to obtain some light variability. In this case, the
variability occurs because the star is occulted by a geometrically thin
magnetospheric disk that moves across the visible surface. This leads to
strong warps in the light curve, which appear, for example, in $\sigma$~Ori~E.

On the other hand, for nonaxisymmetric fields, the variability appears even when
the magnetic and rotational axis coincide. In this case, the disk geometrically
flares and warps in the light curve are caused by the occultation of
a large part of the stellar surface by the flaring disk. The occulting matter is
typically optically thin, which leads to weak dips such as that observed in
HD~37776. 

A combination of low-order axisymmetric multipole with weak higher-order
nonaxisymmetric multipoles leads to the warping of the originally symmetric equilibrium
structure. This could appear even in typical magnetic, hot stars, which are
dominated by a dipolar field. This combination of axisymmetric and
non-axisymmetric multipoles might introduce structures that can explain
the tiny features observed in the light curves of chemically peculiar stars.

\begin{acknowledgements}
The authors thank Profs.~O.~Kochukhov and S.~P.~Owocki for the discussion
of the topic and an anonymous referee for valuable comments which
helped us to significantly improve the paper.
\end{acknowledgements}

\bibliographystyle{aa}
\bibliography{papers}

\end{document}